\documentclass[aps,prl,twocolumn,superscriptaddress]{revtex4-2}
\usepackage{tikz}
\usepackage[english]{babel}
\usepackage[utf8]{inputenc}
\usepackage{indentfirst}
\usepackage{amsmath}
\usepackage{amssymb}
\usepackage{eufrak}
\usepackage{physics}

\newcommand{\La}{\mathcal{L}}

\newcommand{\AAA}{\mathcal{A}}
\newcommand{\TT}{\mathcal{T}}
\newcommand{\BB}{\mathcal{B}}

\newcommand{\CC}{\mathcal{C}}
\newcommand{\DD}{\mathcal{D}}

\newcommand{\complex}{\mathbb{C}}

\newcommand{\eps}{\varepsilon}

\usepackage{amsthm}

\newcommand{\secc}[1]{\textit{#1}.---}

\makeatletter
\renewcommand*\env@matrix[1][*\c@MaxMatrixCols c]{%
  \hskip -\arraycolsep
  \let\@ifnextchar\new@ifnextchar
  \array{#1}}
\makeatother

\usepackage{ifpdf}

\ifpdf
\usepackage{epstopdf}
\usepackage[pdftex,colorlinks,urlcolor=blue,citecolor=blue,linkcolor=blue]{hyperref}
\else
\usepackage[hypertex,colorlinks,urlcolor=blue,citecolor=blue,linkcolor=blue]{hyperref}
\fi
\newcommand{\elte}{\affiliation{MTA-ELTE “Momentum” Integrable Quantum Dynamics Research Group, Department of
    Theoretical Physics, E\"otv\"os Lor\'and University, Budapest, Hungary}}
\newcommand{\bme}{\affiliation{Department of Theoretical Physics, Budapest University of Technology and
      Economics, Budapest, Hungary}}

\begin{document}

\title{
Matrix product symmetries and breakdown of thermalization from hard rod deformations
}
\author{M\'arton Borsi}
\elte
\author{Levente Pristy\'ak}
\bme
\elte
\author{Bal\'azs Pozsgay}
\elte

\begin{abstract}
We construct families of exotic spin-1/2 chains using a procedure called ``hard
rod deformation''. We treat both integrable and non-integrable examples. The models 
possess a large
non-commutative symmetry algebra, which is generated by matrix product operators with fixed small bond dimension.
The symmetries lead to Hilbert space fragmentation and to the breakdown of thermalization.
As an effect, the 
models support persistent oscillations in non-equilibrium situations. 
Similar symmetries have been reported earlier in integrable models, but here we show that they
also occur in non-integrable cases.
\end{abstract}

\maketitle

\secc{Introduction} Symmetry is a key concept in physics. The search for fundamental theories of nature is guided by
symmetry principles, and by Noether's theorem every continuous symmetry of a model leads to a conservation
law. Therefore, it is of fundamental importance to understand: What kind of symmetries can exist in a
certain physical model?
In quantum spin systems on the lattice typical symmetries are those that follow from
the geometrical arrangements (translations and rotations of the lattice), and from ``internal'' symmetries (spin
reflections or rotations). However, in recent years there is growing interest to explore various types of generalized
symmetries that can exist in condensed matter systems (see for example \cite{generalized-symmetries}).

Symmetries crucially affect the dynamical properties of many-body systems.
Unconventional symmetries typically lead to the breakdown of ergodicity, and they can enhance or sometimes diminish
the transport processes in the system. A famous example is provided by integrable models, which possess an infinite set
of extra conservation laws 
\cite{sutherland-book,Korepin-Book}. As an effect, such
models equilibrate to states described by the Generalized Gibbs Ensemble \cite{rigol-gge,JS-CGGE} and they support
ballistic transport \cite{doyon-ghd,jacopo-ghd}.

Other examples for unconventional symmetries are seen in  models with Hilbert space fragmentation
\cite{tibor-fragment,fragment-fracton-2,hfrag-review,fragmentation-scars-review-2,fragmentation-alternative,maksym-east}. In
these  
models there is an exponentially growing number of kinetically disconnected sectors in the Hilbert space. In parallel,
families of such models have a symmetry algebra whose dimension also grows exponentially with the volume
\cite{fragm-commutant-1}. These extra symmetries lead to the breakdown of ergodicity and to
the slowdown of transport (for classical counterparts of this phenomenon see \cite{KCM-review1,KCM-review2}). 
Fragmented models are typically non-integrable, but integrable examples are also known
\cite{fragm-commutant-1,sajat-folded,sajat-hardrod}. 

In this work we consider a specific mechanism for Hilbert space fragmentation, which allows for unconventional
symmetries with striking consequences for non-equilibrium dynamics. Our examples are spin-1/2 chains which are
obtained from ``hard rod deformation'' of short range Hamiltonians. The symmetries of the final Hamiltonians include
standard local $U(1)$-symmetries, but also a large family of unconventional symmetries represented by
Matrix Product Operators (MPO) with 
small bond dimension. Our models are generally non-integrable, and the algebra of the MPO symmetries is not
commutative.

\secc{Models} Our main models are spin-1/2 chains.
The local basis states are denoted as
$\ket{\uparrow}$, $\ket{\downarrow}$, and we use the short notations $X_j, Y_j, Z_j$ for the Pauli matrices acting on
site $j$. We use the local projectors $P_j=(1+Z_{j})/2$ and $N_j=(1-Z_{j+1})/2$, and also the two-site projectors
$\Pi_{j,k}^\pm=(1\pm Z_jZ_k)/2$. In all cases we treat
extensive and translationally invariant Hamiltonians defined as $H=\sum_{j=1}^L h(j)$, with an operator density $h(j)$
which is a short range 
operator. We work with periodic boundary conditions. 

We treat a family of models defined by
\begin{equation}
  \label{HHH}
  h(j)=h_{f}(j)+\Delta h_{ZZ}(j)+\kappa h_{ni}(j).
\end{equation}
The first term is the kinetic part of the Hamiltonian, describing controlled hopping:
\begin{equation}
  \label{f}
  h_{f}(j)=(X_{j+1}X_{j+2}+Y_{j+1}Y_{j+2}) \Pi^+_{j,j+3}.
\end{equation}
The other two terms describe interactions, they are diagonal in the given basis, and
they span 4 and 6 sites, respectively: 
\begin{equation}
  \begin{split}
    h_{ZZ}(j)&=\Pi^+_{j,j+3}\Pi^{-}_{j+1,j+2},\\
    h_{ni}(j)&=\Pi^+_{j,j+5} \Pi^-_{j+1,j+2}\Pi^-_{j+2,j+3}\Pi^-_{j+3,j+4}.
  \end{split}
\end{equation}

Without interactions  ($\Delta=\kappa=0$) we have the {\it folded XXZ model},
which describes the high temperature dynamics of the
XXZ Heisenberg spin chain in the large anisotropy limit
\cite{folded1,folded2,sajat-folded,folded-jammed,folded4}. The folded XXZ model  also appeared in \cite{folded0}, and it
is closely related to stochastic models treated in \cite{folded-XXZ-stochastic-1}. 
It is an integrable model, which can be solved exactly by the Bethe Ansatz \cite{folded1,sajat-folded}.

Switching on $\Delta\ne 0$ but keeping $\kappa=0$ we obtain the so-called {\it hard rod deformed XXZ
model} introduced in \cite{sajat-hardrod}. It is also an integrable model, which is closely related to the actual XXZ
model and also to the constrained models of \cite{constrained1,constrained2,constrained3}. Switching on $\kappa\ne 0$
breaks integrability \footnote{Supplemental  Materials to ``Matrix product symmetries and breakdown of thermalization
  from hard rod deformations''}. 

\secc{Dynamics}
The kinetic term in \eqref{f} generates the transitions 
$\ket{\uparrow\uparrow\downarrow\uparrow}\leftrightarrow \ket{\uparrow\downarrow\uparrow\uparrow}$ and
$\ket{\downarrow\uparrow\downarrow\downarrow}\leftrightarrow \ket{\downarrow\downarrow\uparrow\downarrow}$
on four sites.
As an effect, single down/up spins can propagate freely in a background of up/down spins, respectively. On the other
hand, states with  isolated domain walls are frozen. For example, the kinetic term acts as zero
on the local configuration
$\ket{\uparrow\uparrow\downarrow\downarrow}$. It follows, that any state which consists only of domains (sequences of
spins with the same orientation, being longer than 2) are frozen.
However, non-trivial dynamics arises when a single particle scatters on an isolated domain wall. In such a case, we observe particle-hole
transmutation: when an incoming particle (for example a down spin in a background of up spins) meets a domain wall, it
continues its path as a hole (in this case, as an up spin in a background of down spins). As a by-product, the domain
wall gets displaced by two sites. This dynamical phenomenon was treated in detail in \cite{quantum-bowling,sajat-folded,vasseur-anom}.

It follows from the structure of the kinetic term and the interaction terms, that the following two $U(1)$ charges are
conserved for arbitrary $\Delta$ and $\kappa$:
\begin{equation}
  \begin{split}
    Q_1=\sum_j Z_j,\qquad
    Q_2=\sum_j Z_j Z_{j+1}.
  \end{split}
\end{equation}
Here $Q_1$ is the global magnetization, while $Q_2$ is (up to normalization) the ``domain wall number''. 

\secc{Matrix Product symmetries} Below we show that our model possesses exotic symmetries for generic values of
the coupling 
constants, also in the non-integrable case. These symmetries are represented by Matrix Product Operators (MPO's), which
commute with the Hamiltonian. 
An MPO is a one-dimensional tensor network where each tensor has two external indices (corresponding to the physical
spaces) and two 
internal indices (corresponding to an auxiliary space $V_a=\complex^D$ with an appropriate constant $D\ge
2$).

We introduce the elementary tensor as a linear operator $\La_{a,j}$ which acts on the tensor product
space $V_a\otimes V_j$, where $V_j=\complex^2$ is the physical space at site $j$. The MPO with periodic boundary
conditions is then defined as
\begin{equation}
  \label{MPOdef}
  \TT=\text{Tr}_a \big[\La_{a,L}\dots \La_{a,2}\La_{a,1}\big].
\end{equation}
We say that $\TT$ is an MPO symmetry if it commutes with the Hamiltonian in every volume $L$.

A distinguishing property of an MPO is that its operator space entanglement entropy
\cite{entangling-power,prosen-op-ent}
is bounded from above by $2\log(D)$. Therefore it satisfies the ``area law'' of entanglement \cite{jerome-op-ent}. In
the special 
case of $\La_{j,a}=o_j$ with $o$ being a one-site operator the MPO becomes proportional to a product
operator. Therefore, {\it an MPO symmetry can be seen as a generalization of strictly local internal symmetries.}

MPO symmetries are known to exist in integrable spin chains with local interactions
\cite{faddeev-how-aba-works,Korepin-Book}. In those models the tensor $\La$ is called the Lax operator, and it depends
on a complex variable (spectral parameter) and possibly some discrete variables too. The resulting
MPO's are called transfer matrices, and they form a commuting family. Extensive conserved charges with short range
operator densities are derived from such families of MPO's.

In our family of models  commuting transfer matrices have been found in
 the integrable cases  in
\cite{sajat-folded,sajat-hardrod}. They fit
into the canonical framework of Yang-Baxter integrable spin chains \cite{faddeev-how-aba-works}, 
generalized to spin chains with medium range interaction \cite{sajat-medium}.
However, those symmetries get broken after switching on $\kappa\ne 0$.

In contrast, we derive new MPO symmetries that hold
in both the integrable and non-integrable cases.
In order to derive these symmetries first we perform a sequence of
transformations on our models.

\secc{Bond model} Following \cite{folded1,sajat-folded} we perform a so-called 
bond-site transformation, which we define on the level of the basis states in the computational basis. The idea is to put
variables on the bonds between the sites, such that the values $\pm 1$ on the bond represent whether the two
neighbouring spins have the same or different values. The original Hamiltonian is invariant with respect to global spin
reflection, therefore it will generate local dynamics for the bond variables. Basis states in the bond model will be
denoted as $\ket{\circ}$ (empty site or spin up) and $\ket{\bullet}$ (occupied site or spin down), which correspond to
identical and opposite spins on neighbouring sites of the original model, respectively. For more details about the
transformation see \cite{Note1}. 

In the bond model the non-zero kinetic transitions are
 $\ket{\circ\bullet\bullet} \leftrightarrow \ket{\bullet\bullet\circ}$.
These are interpreted as a one site translation of dimers or ``hard rods'', which are particles spanning two sites. They are the
mobile particles in these models, and they motivated the use of the expression ``hard rod deformation'' \cite{sajat-hardrod}.
In contrast, single $\ket{\bullet}$ states embedded in a vacuum of empty sites
are immobile on their own. They are displaced when a mobile particle scatters
on them.

\secc{The XXC models} The bond models  can be mapped further to spin chains with three dimensional local
spaces. This mapping is non-local and volume changing, and it appeared among others in \cite{sajat-folded} but also much
earlier in \cite{folded-XXZ-stochastic-1} (see also \cite{tracer-dynamics}). The mapping
is defined as follows.

In the computational basis the states can be seen as a sequence of $\circ$ and $\bullet$ ``characters'', and these
sequences are  
translated into sequences consisting of the numbers 0, 1 and 2. The original
sequence is ``read'' from the left to the right. If a $\circ$ is encountered then one writes down a 1. If a $\bullet$
is encountered, then one also reads the next character. In case of a $\bullet$ or $\circ$ one writes down a 0 or a 2,
respectively. This gives the local transformation rules
\begin{equation}
  \label{bond2xxc}
  \ket{\circ}\to \ket{1},\quad \ket{\bullet\bullet}\to\ket{0},\quad\ket{\bullet\circ}\to\ket{2}.
\end{equation}
This mapping is volume changing: the length of the new sequence depends on the content of the original
sequence. This implies that different sectors of the Hilbert space of the original model will be mapped to Hilbert
spaces of the new spin chain with varying lengths. 

The transformation  induces a mapping for the Hamiltonians. The transformation of the basis states is
strongly non-local, therefore locality is typically lost on the level of the Hamiltonians. Nevertheless it is possible
to select certain local Hamiltonians which remain local after the mapping
\cite{sajat-folded,sajat-hardrod,folded-XXZ-stochastic-1,tracer-dynamics}, and our family of models also has this property.

We introduce notations for operators acting on the three dimensional local spaces. We have
$s^-_{\alpha}=\ket{\alpha}\bra{0}$ with $\alpha=1, 2$ and they can be arranged into a two-dimensional vector ${\bf
  s^-}=(s^-_1,s^-_2)$. Furthermore ${\bf s^+}=({\bf s^-})^\dagger$, and we also introduce the projectors
$n=\ketbra{0}{0}$, $p=\ketbra{1}{1}+\ketbra{2}{2}$.  
Direct computation shows \cite{Note1} that our model Hamiltonians are eventually mapped to short range Hamiltonians with density
\begin{equation}
  \label{HXXC}
   h^C(j)=h^C_{f}(j)+\Delta h^C_{ZZ}(j)+\kappa h^C_{ni}(j),
 \end{equation}
 with
 \begin{equation}
   \begin{split}
   h^C_{f}(j)&=   {\bf s^-}_{j}\cdot {\bf s^+}_{j+1} +  {\bf s^+}_{j}\cdot {\bf s^-}_{j+1},\\
     h^C_{ZZ}(j)&=n_j p_{j+1}+p_j n_{j+1},\\
     h^C_{ni}(j)&= n_{j} n_{j+1} p_{j+2}+p_j n_{j+1} n_{j+2}.
   \end{split}
\end{equation}
The model with $\Delta=\kappa=0$ appeared in \cite{su3-xx} and it is closely related to the strong coupling limit of the
Hubbard model (also known as the $t-0$ model). The model with $\Delta\ne 0$ but $\kappa=0$ appeared in \cite{XXC} (and
in a special case in 
\cite{folded-XXZ-stochastic-1}) and it was called the XXC model. The non-integrable perturbation appears to be new; 
we call it the deformed XXC model.

\secc{Spin-charge separation} The kinetic terms in \eqref{HXXC} generate the transitions
 $\ket{01}\leftrightarrow\ket{10},\ \ket{02}\leftrightarrow\ket{20}$.
The transition $\ket{12}\leftrightarrow\ket{21}$ is forbidden, thus the relative ordering of the basis states $\ket{1}$
is $\ket{2}$ can not be changed during time evolution.

We can regard the local state $\ket{0}$ as the vacuum, and the states $\ket{1}$ and $\ket{2}$ as a particle (charge) with an
internal degree of freedom (spin). Then we can perform a spin-charge separation: we specify each basis state by giving
the positions and the spins of the particles. The Hamiltonians are such that {\it the spin-charge separation leads to 
  exactly decoupled dynamics, the spin part of the wave function is a constant of motion, and it does not influence
  the motion of the particles.} This is a non-trivial
property, which we prove in detail in \cite{Note1}.

This phenomenon was
already observed in a number of works dealing with similar models
\cite{folded-XXZ-stochastic-1,bruno-hubbard,sajat-folded,tracer-dynamics,prosen-anomalous-universal}.
It induces Hilbert space fragmentation: different values of the spin pattern  all correspond to
different irreducible sectors in the Hilbert space. This phenomenon underlies the existence of the exotic
symmetries of all our models. Furthermore, it allows for exact solutions of real time dynamics in similar models \cite{prosen-MM1,bruno-hubbard,prosen-anomalous-universal,sasha-anyon,sajat-t0-diff}.

\secc{Symmetries for XXC} We construct MPO symmetries for the deformed XXC models, and afterwards
we generalize the construction for our original family \eqref{HHH}.

In the deformed XXC models we construct MPO's with fixed bond dimension 2. The key idea is that the MPO's should act
only on the spin degrees of freedom, while leaving the particle positions intact \cite{folded-XXZ-stochastic-1}. This
will guarantee that the MPO's commute with the Hamiltonian.
Generally such operators are very
non-local, but there exist representatives with the desired MPO structure. The majority of our results for the MPO's is new.

In the XXC case we choose the local tensor $\La$ as
\begin{equation}
  \label{AdefXXC2}
  \La= I\otimes \ketbra{0}{0}+\sum_{\alpha,\beta=1,2} F^{(\alpha,\beta)}\otimes \ketbra{\alpha}{\beta}.
\end{equation}
Here $I$ and $F^{(\alpha,\beta)}$ are five matrices of size $2\times 2$ which act on the auxiliary space, and specifically  $I$ is the
identity matrix. Such an MPO acts as the identity on every local vacuum state $\ket{0}$, but typically it has a
non-trivial action on the spin degrees of freedom. The resulting MPO's do not change the position of the particles, but
they can modify the spin pattern.

We consider two sub-classes of such MPO's. In one class the resulting MPO's are diagonal, which can be achieved by
setting $F_{12}=F_{21}=0$. Such MPO's do not change the spin pattern, but their eigenvalues  (diagonal matrix elements)
do depend on it. These MPO's all commute with each other and also the Hamiltonian.

The number of independent parameters of these MPO's can be reduced to 5, and representatives can be chosen for example
as 
\begin{equation}
  \label{F1122}
  F^{(1,1)}=
  \begin{pmatrix}
    x & y \\ y & z
  \end{pmatrix},\quad
 F^{(2,2)}=
  \begin{pmatrix}
    u & 0 \\ 0 & v
  \end{pmatrix}.
\end{equation}

The second class of MPO symmetries changes the spin pattern. We concentrate on those MPO's which conserve the number of
$\ket{1}$ and $\ket{2}$ states.
This can be achieved by the following matrices with  5 independent parameters:
\begin{equation}
  \label{offd}
  \begin{split}
&    F^{(1,2)}=(F^{(2,1)})^\dagger=\gamma\sigma^-, \\
    &\hspace{1cm}F^{(1,1)}=
  \begin{pmatrix}
    \alpha & \\ & \delta \\
  \end{pmatrix},\
  F^{(2,2)}=
    \begin{pmatrix}
    \beta & \\ & \eps \\
  \end{pmatrix}.
  \end{split}
\end{equation}
Our diagonal MPO's are included in the results of \cite{fragm-commutant-1}, but
the off-diagonal ones appear to be new. 

\secc{Main results} Now we pull back these MPO symmetries to the original family of models given by \eqref{HHH}. This is
a non-trivial task, because the transformation rule \eqref{bond2xxc} causes strong non-locality. However, the action of
the two
classes of MPO's that we introduced can be 
emulated by an MPO with fixed bond dimension even in the original model. The auxiliary dimension needs to be enlarged in
order to deal with the non-local effects, but afterwards it will not depend on the volume. It is important that {\it the
transformation between the models can not be described by an MPO with fixed bond dimension}: This happens only for the
selected symmetry operators that we construct.

In order to find the actual MPO's, we use the techniques discussed in
\cite{mps-automata}. We view the MPO as an ``automaton'' with a finite number of internal states, which are changed as
the MPO acts on the physical spin chain. These internal states and their transitions will encode the rules  \eqref{bond2xxc}
and also the bond-site transformation.

We construct two families of MPO symmetries which we denote as $\TT^d$ and $\TT^{o}$,
corresponding to the diagonal and off-diagonal classes above. In both cases we
expand $\La$ as
\begin{equation}
  \label{MPOrep}
  \La=\AAA\otimes\ketbra{\uparrow}{\uparrow}+
  \BB\otimes\ketbra{\uparrow}{\downarrow}+
 \CC\otimes\ketbra{\downarrow}{\uparrow}+
  \DD\otimes\ketbra{\downarrow}{\downarrow},
\end{equation}
where $\AAA, \BB, \CC, \DD$ are sparse matrices of size $D\times D$ acting on the auxiliary space.

For the family $\TT^d$ the auxiliary space has dimension $D=8$ and we view it as the tensor product
$\complex^2\otimes\complex^2\otimes\complex^2$. The MPO's depend on 5 independent
parameters, and they are diagonal, which is ensured by $\BB=\CC=0$. They commute with each
other  and also with the local charges $Q_1$ and $Q_2$. The concrete matrices are
\begin{equation}
  \begin{split}
    \AAA= N \otimes F^{(1,1)} \otimes P &+ \sigma^+ \otimes F^{(2,2)} \otimes \sigma^- + \\
    N \otimes &I \otimes \sigma^- + P \otimes I \otimes \sigma^+, \\
    \DD= P \otimes F^{(1,1)} \otimes P &+ \sigma^- \otimes F^{(2,2)} \otimes \sigma^- + \\
    P \otimes &I \otimes \sigma^- + N \otimes I \otimes \sigma^+,\\    
  \end{split}
\end{equation}
where $N$ and $P$ are  projectors introduced above, and $F^{(1,1)}$ and $F^{(2,2)}$ are given in \eqref{F1122}.

In the case of the family $\TT^o$ the auxiliary space has dimension $D=10$, and $\La$ depends on 5 independent
parameters. These MPO's are generally not diagonal, and they do not commute with each other.
Concrete matrix elements are \cite{Note1}
  \begin{equation}
    \begin{split}
  	\mathcal{A}_{1,2} &= \mathcal{A}_{3,4} = \mathcal{A}_{5,4} = \mathcal{A}_{7,6} = \mathcal{A}_{9,10} = 1,\\
  	\mathcal{D}_{2,1} &= \mathcal{D}_{4,5} = \mathcal{D}_{6,7} = \mathcal{D}_{8,9} = \mathcal{D}_{10,9} = 1, \\
  	\mathcal{A}_{6,6} &= \mathcal{D}_{1,1} = \alpha,\quad  \mathcal{A}_{2,6} = \mathcal{D}_{7,1} = \beta,  \\
    	\mathcal{A}_{4,6} &= \mathcal{B}_{1,3} = \mathcal{C}_{6,8} = \mathcal{D}_{9,1} = \gamma, \\
    	\mathcal{B}_{3,3} &= \mathcal{B}_{5,3} = \mathcal{C}_{8,8} = \mathcal{C}_{10,8} = \delta, \\
    	\mathcal{B}_{9,3} &= \mathcal{C}_{4,8} = \varepsilon.      
   \end{split}
  \end{equation}
These MPO's commute with $Q_2$, because they originate from the MPO's given by \eqref{offd}, which conserve the ``spin'' in
the XXC models, eventually leading to conservation of the number of domain walls in the original models. However, they
break the global magnetization $Q_1$, because {\it they generate a displacement of the domain walls}. 
  
The MPO's do not depend on the parameters $\Delta, \kappa$: they are symmetries for the full
family of models. The diagonal MPO's commute with the Hamiltonian densities $h(j)$ separately: they belong to
the commutant algebra \cite{fragm-commutant-1}. The off-diagonal
ones commute only with the full Hamiltonian.

\secc{Persistent oscillations} We explore the dynamical consequences of the MPO symmetries. We consider real time
evolution started from a selected initial state
\begin{equation}
  \ket{\Psi_0}=\otimes_{j=1}^L (\ket{\uparrow}+\ket{\downarrow})/\sqrt{2},
\end{equation}
which is a state completely polarized in the $x$-direction. This state breaks the $U(1)$-invariance associated with the
global magnetization. We consider time evolution generated by $H+h Q_1$, where $H$ is given by \eqref{HHH} with generic
values of the coupling constants and $h$ is a magnetic field. We focus on the time evolution of the local observable
$X_j$; for simplicity we will drop the site index $j$ in the notation.

\begin{figure}[t]
  \centering
  \includegraphics[scale=0.3]{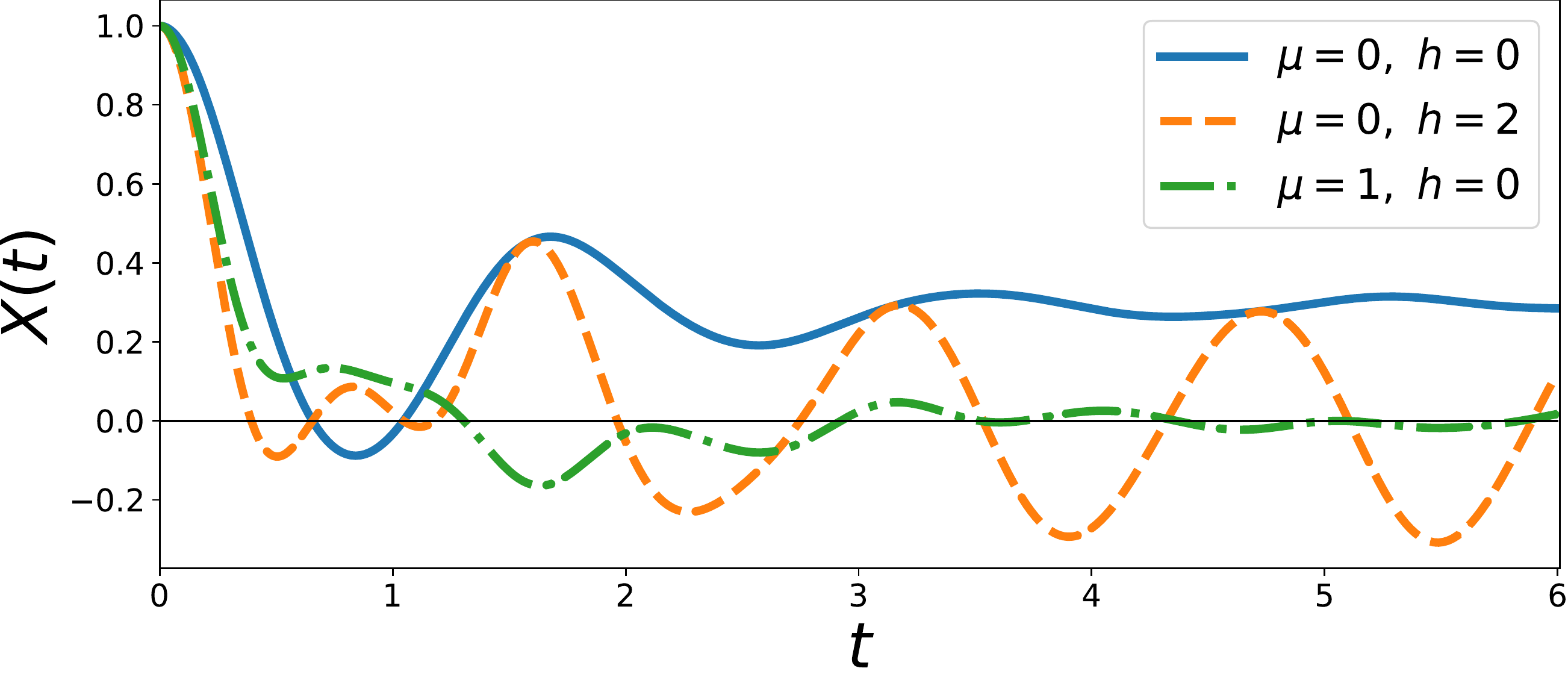}
  \caption{Real time dynamics from a selected initial state, with a non-integrable Hamiltonian with $\Delta=0.2$
    and 
    $\kappa=0.5$. 
    The second curve is obtained after adding a magnetic field $h$.
    The third curve is obtained after adding a perturbing term $H'$ which
    breaks the MPO symmetries.
  }
  \label{fig}
\end{figure}

We performed the numerical computation of the real time evolution using the iTEBD method
\cite{vidal-itebd0,vidal-itebd1a}. Our data is 
 presented on Figure \ref{fig}, for details see \cite{Note1}.

The local operator $X$ breaks the $U(1)$-symmetry generated by $Q_1$. In the absence of extra symmetries it is expected
that the mean value $\bra{\Psi_0}{X(t)}\ket{\Psi_0}$ drops to zero in the long time limit, for both integrable and non-integrable
cases. However, in our case we observe that $X(t)$ has a non-zero stationary value for $h=0$, and for $h\ne0$ it shows
non-decaying oscillations with  frequency $h$. The reason for this
phenomenon is that the off-diagonal MPO symmetries also break the given $U(1)$ charge. Adding one more perturbation
$H'=\mu \sum_j Z_j Z_{j+1}Z_{j+2}$ breaks all MPO symmetries, and in this case we observe relaxation to zero, as expected.

Persistent oscillations were reported earlier in relation with integrability \cite{time-crystal,sajat-folded}
and also in models with quantum scars \cite{pxp,hfrag-review,fragmentation-scars-review-2}.
The novelty of the present results is that we find the same effects in non-integrable models, explained by the MPO symmetries.
In our models ergodicity breaking extends over essentially the full
Hilbert space, therefore the phenomenon is not related to quantum scars.

\secc{Discussion} Our mechanism for Hilbert space fragmentation allows for unusual MPO
symmetries, which hold in the integrable and non-integrable cases too. The MPO symmetries generate a non-commutative
algebra, therefore the models should be seen as having  quantum fragmentation.
In the literature there have been few examples for quantum fragmentation
\cite{fragm-commutant-1,read-saleur,maksym-east}, and our models provide a new mechanism for this. Also, they appear to
be the 
first examples of non-integrable models with off-diagonal MPO symmetries.

Our models have {\it strong fragmentation} \cite{tibor-fragment}, because the symmetries affect the whole spectrum in a
non-trivial manner. This leads to exponentially large degeneracies for almost all states, but the concrete degeneracies depend
on the state \cite{sajat-folded,sajat-hardrod}.
Our symmetry operators are similar in
essence to the ``statistically localized integrals of motion'' found in \cite{tibor-fragm-SLIOM}, but it is a novel
result that we construct them in the form of MPO's with low bond dimension.

The XXC models that appeared in our study have the special property that spin-charge separation is exact, the spin
pattern is always conserved, and it does not influence the charge degrees of motion. It was argued in
\cite{prosen-anomalous-universal} that in such models the spin transport 
has anomalous fluctuations. This is believed to be true also for the folded XXZ model
\cite{vasseur-anom,prosen-anomalous-universal}, although the rigorous proofs of \cite{prosen-anomalous-universal}  do
not apply in that case. We conjecture that our family of models also displays anomalous fluctuations, in both the
integrable and non-integrable cases.

\begin{acknowledgments}
  We are thankful to Frank G\"ohmann, Enej Ilievski, Sanjay Moudgalya, Tibor Rakovszky and Lenart
Zadnik  for useful discussions. 
\end{acknowledgments}



%

\widetext
\newpage

\begin{center}
  \textbf{\large Supplemental Materials:}
  \medskip

  \textbf{\large ``Matrix product symmetries and breakdown of thermalization from
  hard rod deformations''}
\end{center}
\setcounter{equation}{0}
\setcounter{figure}{0}
\setcounter{table}{0}
\setcounter{page}{1}
\makeatletter
\renewcommand{\theequation}{S\arabic{equation}}
\renewcommand{\thefigure}{S\arabic{figure}}
\renewcommand{\bibnumfmt}[1]{[S#1]}
\renewcommand{\citenumfont}[1]{S#1}

\section{Content}

In these Supplemental Materials we provide a number of technical details for our results in the main text:

\begin{itemize}
\item We treat the bond-site transformation in more detail.
\item We consider the transformation to the deformed XXC models and show that our local Hamiltonians keep their locality
  under the mapping.
\item We establish the spin-charge separation in the XXC models in more detail. We also provide the local Hamiltonian
  that dictates the motion of the charge part of the wave functions. Furthermore, we demonstrate that one of the
  interaction terms indeed breaks the integrability of the models.
\item We provide technical details for the derivation of the MPO symmetries in our original spin-1/2 models.
\item We give some details about our numerical procedures. 
\end{itemize}

\section{The bond-site transformation}

\label{suppm:bst}

In this Section we provide more details for the bond-site transformation. As it is written in the main text, the idea is
to put new variables on the bonds between the sites of the original spin chain. The value of the bond variable is $\pm
1$ depending on whether the two neighbours have the same or different orientations. In our notations of the basis elements:
\begin{equation}
  \ket{\uparrow\uparrow},\   \ket{\downarrow\downarrow}\to \ket{\circ},\qquad \qquad
  \ket{\uparrow\downarrow},\   \ket{\downarrow\uparrow}\to \ket{\bullet}.
\end{equation}

On the operator level this definition gives the transformation rule
\begin{equation}
\label{Zrule}
  Z_{j}Z_{j+1}\quad\to\quad Z_{j+1/2},
\end{equation}
where the half-shift signals that the new variable (or operator) is defined on the bonds. 

A spin-flip on a single site of the original chain implies a
change in two neighbouring bond variables, leading to the transformation rule
\begin{equation}
  \label{Xrule}
  X_{j}\quad\to\quad X_{j-1/2}X_{j+1/2}.
\end{equation}
The rules \eqref{Zrule}-\eqref{Xrule} are sufficient to formally define the transformation on an operatorial
level.

Strictly speaking the transformation is defined only for a chain with free boundary conditions or in the half-infinite
limit.
Nevertheless, the rules \eqref{Zrule}-\eqref{Xrule} uniquely define a mapping for Hamiltonians which are spin reflection
symmetric. 

The various terms of our family of models can be transformed using the steps
\begin{equation}
  \begin{split}
    X_jX_{j+1}&\quad\to\quad X_{j-1/2}X_{j+1/2}X_{j+1/2}X_{j+3/2}=X_{j-1/2}X_{j+3/2},\\
Y_jY_{j+1}=-Z_jZ_{j+1}X_jX_{j+1}&\quad\to\quad - X_{j-1/2}Z_{j+1/2}X_{j+3/2},\\
Z_{j}Z_{j+3}=Z_j Z_{j+1} Z_{j+1}Z_{j+2}Z_{j+2}Z_{j+3}&\quad\to\quad Z_{j+1/2} Z_{j+3/2}Z_{j+5/2}.\\
  \end{split}
\end{equation}
Combining these formulas we obtain that the original family of Hamiltonians is mapped to  the bond model with
Hamiltonian density
\begin{equation}
  \label{Hbond}
   h^B(j)=h^B_{f}(j)+\Delta h^B_{ZZ}(j)+\kappa h^B_{ni}(j),
 \end{equation}
with
\begin{equation}
  \label{bondd}
  \begin{split}
    h^B_{f}(j)&=(X_j X_{j+2}+Y_j Y_{j+2})P_{j+1},  \\
h^B_{ZZ}(j)&= (N_jP_{j+2}+P_jN_{j+2}) P_{j+1},\\
h^B_{ni}(j)&= (N_jP_{j+4}+P_jN_{j+4}) P_{j+1}P_{j+2}P_{j+3}.
  \end{split}
\end{equation}
Here we deleted the half shifts.

Strictly speaking, the original and the bond models are identical to each other only up to boundary effects. In the case
of free boundary 
conditions the original model in  volume $L$ is equivalent to the bond model in a volume $L-1$. In case of periodic
boundary conditions, the volume is kept the same, and the two models are identical only in that sector of the bond model,
which has an even number of $\ket{\bullet}$ states. 

\section{Mapping to XXC}

\label{sec:xxc}

The mapping to the deformed XXC models can be proven via direct computation in the real space basis.
The mapping is performed sequentially in the computational basis  and the fundamental rules are
\begin{equation}
  \label{bond2xxc2}
  \ket{\circ}\to \ket{1},\quad \ket{\bullet\bullet}\to\ket{0},\quad\ket{\bullet\circ}\to\ket{2}.
\end{equation}
This mapping is  non-local because it is volume changing: the length of the new sequence depends on the content of the
original state. Furthermore,
the interpretation of a certain part of the sequence depends on the previous values in the sequence as well. Consider
the following two examples for a mapping in a volume 5:
\begin{equation}
  \begin{split}
    \ket{\circ\circ\bullet\bullet\circ}\quad&\to\quad \ket{1101},\\
       \ket{\circ\bullet\bullet\bullet\circ}\quad&\to\quad \ket{102}.\\
  \end{split}
\end{equation}
The difference is not only in the length of the resulting sequences, but also in the interpretation of the two
$\ket{\bullet}$ states on positions 3 and 4 from the left: In the first case, they are interpreted as the composite state $\ket{0}$,
whereas in the second case they become parts of the two different composite states $\ket{0}$ and $\ket{2}$. This shows
that the mapping is indeed very non-local, and the concrete interpretation of certain segments can not be viewed in an
isolated way.

Nevertheless, it is possible to perform a mapping between local Hamiltonians, by a careful computation of all the
possible matrix elements. Starting with the kinetic term in the bond model, we observe the transition matrix elements
\begin{equation}
 \ket{\circ\bullet\bullet} \leftrightarrow \ket{\bullet\bullet\circ}.
\end{equation}
The key observation is that these transitions will be mapped to the transitions
\begin{equation}
 \ket{10}\leftrightarrow\ket{01}\quad\text{or}\quad\ket{20}\leftrightarrow\ket{02},
\end{equation}
depending on the state of the neighbouring sites. A careful case by case analysis shows that each transition matrix
element in the original model is mapped to a transition matrix element in the XXC model with the same multiplicity. This
gives the mapping
\begin{equation}
  \sum_j h^B_{f}(j)\quad\to\quad \sum_j h^C_{f}(j).
\end{equation}
It is important that the mapping does not work for individual operator densities, due to the non-local effects. However, it
works for the total Hamiltonian, up to boundary effects. This mapping for the kinetic terms appeared already in
\cite{folded-XXZ-stochastic-1s,sajat-foldeds}, and a closely related mapping was derived in \cite{tracer-dynamicss}.

The mapping of the diagonal interaction terms needs to be investigated separately. Consider first the term $h^B_{ZZ}(j)$
in the bond model, given in \eqref{bondd}. 
This term gives an eigenvalue 1 to the sequences $\ket{\circ\bullet\bullet}$ and
$\ket{\bullet\bullet\circ}$, wherever  they are along the chain. Let us introduce a definition: We say that a certain segment
of the chain is an ``island'' if it consists only of $\ket{\bullet}$ states and the immediate neighbours of the given
segment are $\ket{\circ}$ states. Then it is easy to see that the operator $\sum_j h^B_{ZZ}(j)$ will give the number of
islands which are at least 2 sites long, multiplied by a factor of  two, corresponding to the 
 two boundaries of each island. The rules \eqref{bond2xxc2}  map each island to a sequence consisting of $\ket{0}$
states, with the boundaries given by either $\ket{1}$ or $\ket{2}$. The term $h^B_{ZZ}(j)$ is sensitive only to islands
longer than 2 sites, therefore, these islands are always mapped to sequences consisting of at least one $\ket{0}$.  The
number of such islands in the XXC models is then ``measured'' by the term $\sum_j h^C_{ZZ}(j)$, where $h^C_{ZZ}(j)$ is
given in the main text. The mapping
for this interaction term appeared already in \cite{sajat-hardrods}.

In a similar way we obtain a mapping
\begin{equation}
 \sum_j h^B_{ni}(j)\quad\to\quad \sum_j h^C_{ni}(j),
\end{equation}
with $h^C_{ni}(j)$ given in the main text.
In the bond model this term measures the number of islands with length $\ge 4$, whereas in the XXC model they will
correspond to islands of $\ket{0}$ states with length $\ge 2$. 

At present, it is not known generally, which local Hamiltonians are transformed to local Hamiltonians via this
mapping. The three terms that we considered were found on a case by case basis. A trivial generalization is to construct
diagonal interaction terms, which would measure the number of even longer islands. However, we expect that there is
a much larger family of Hamiltonians which will keep the locality under the mapping.

\section{Spin-charge separation}

\label{sec:spincharge}

Here we perform the spin-charge separation in the deformed XXC models on the level of the wave functions. This
separation is at the basis of our constructions for the MPO symmetries.

In the deformed XXC models let us consider a state with $N$ particles in a volume of length $L$. The
state can be described as
\begin{equation}
  \label{statesep}
  \ket{\Psi}=\sum_{x_1<\dots<x_N}\sum_{a_j=1,2}
  \chi(x_1,x_2,\dots,x_N)
\psi_{a_1,a_2,\dots,a_N}
  \prod_{j=1}^{N}s^-_{a_j}(x_j)  \ket{\emptyset},
\end{equation}
where
\begin{equation}
  \ket{\emptyset}=\ket{0000\dots 0}
\end{equation}
is the vacuum state and $s^-_{a_j}(x_j)$ with $a_j=1,2$ are creation operators of particles with ``spin'' $a_j$. In this
formula $\chi(x_1,x_2,\dots,x_N)$ is the wave function describing the charge degrees of freedom (positions of
particles), whereas $\psi_{a_1,a_2,\dots,a_N}$ describes the spin degrees of freedom.

The Hamiltonian generates time evolution for both the charge and the spin part. In a generic model these two equations of
motion are coupled. However, in the special case of the deformed XXC models there is an exact decoupling: the spin part
of the wave function is a constant of motion, and it does not influence the motion of the charge part. This is seen by
considering all possible matrix elements of the Hamiltonian, and observing that both the hopping and the diagonal
interaction terms are completely insensitive to the spin part, and the hopping can never change the order of the
particles.

This property was known for a long time in the large coupling limit of the Hubbard model (also known as the $t-0$
model), see for example \cite{infinite-U-Hubbard-corrs,bruno-hubbards,sasha-anyons,sajat-t0-diffs}. It was also known to
hold in special models with quantum gases \cite{zvonarev-spin-charge-1s} and certain cellular automata
\cite{prosen-MM1s,prosen-MM1bs}. This property enables the computation of spin transport coefficients with semi-classical
methods \cite{tracer-dynamicss}, and it also underlies the presence of anomalous fluctuations
\cite{prosen-MM-anomalouss,prosen-anomalous-universals}.

In our models the charge part of the wave function evolves according to the Hamiltonian (acting on a Hilbert space with
two dimensional local spaces)
\begin{equation}
  \label{chargeH}
  h^{ch}(j)=h_{XX}(j)+\Delta h^{ch}_{ZZ}(j)+\kappa h^{ch}_{ni}(j),
\end{equation}
where
\begin{equation}
  \begin{split}
  h_{XX}(j)&= X_jX_{j+1}+Y_j Y_{j+1},\\
  h^{ch}_{ZZ}(j)&=(N_jP_{j+1}+P_{j}N_{j+1}),\\
    h^{ch}_{ni}(j)&=(N_jP_{j+1}P_{j+2}+P_{j}P_{j+1}N_{j+2}).\\
  \end{split}
\end{equation}
This is seen after ``dropping the spin part'': We construct a wave function for a spin-1/2 chain via
\begin{equation}
  \ket{\Psi}=\sum_{x_1<\dots<x_N}
  \chi(x_1,x_2,\dots,x_N)
  \prod_{j=1}^{N}\sigma^-(x_j)  \ket{\emptyset},
\end{equation}
where now $\ket{\emptyset}=\ket{\uparrow\uparrow\dots\uparrow}$, and observe that the equation of motion for $
\chi(x_1,x_2,\dots,x_N)$ dictated by \eqref{chargeH} is exactly the same as in the XXC models. Note that all terms in
\eqref{chargeH} come directly from the deformed XXC Hamiltonians (eq. (7) and (8) in the main text), after ``dropping
the spin part''. 

Choosing $\kappa=\Delta=0$ the model given by \eqref{chargeH} is the XX model, which is solvable by Jordan-Wigner
transformation. Keeping $\kappa=0$ but switching on a finite $\Delta$ we obtain the XXZ Heisenberg chain (although with
an un-conventional choice for the $\Delta$ anisotropy parameter). Finally, the model with $\kappa\ne 0$ is
non-integrable.
Based on the similarity of the expressions (and its physical content) we say that the bond model Hamiltonian
\eqref{Hbond} should be seen as the ``hard rod deformation'' of the models given by \eqref{chargeH}.

In almost all of our derivations we dismissed the boundary effects which occur at various stages of the
mapping. However, in the case of periodic boundary conditions the problems which arise can be remedied by introducing a
``twist''. Here we just present the main statement, and for a more detailed derivation we refer to \cite{sajat-foldeds},
which treats the mapping from the folded XXZ model to the XXC model with $\kappa=\Delta=0$. 

Let us consider a ``twisted'' kinetic term
\begin{equation}
  \begin{split}
  h_{XX}(j,\theta)&= 2(\sigma^-_j \sigma^+_{j+1} e^{i\theta}+\sigma^+_j \sigma^-_{j+1} e^{-i\theta}),
  \end{split}
\end{equation}
which is meant to replace the kinetic term in \eqref{chargeH}. The real parameter $\theta$ arises from twisted boundary
conditions for the wave functions, but we choose to distribute the twist evenly along the chain. Then we state the
following:

{\bf Proposition:} All eigenvalues of the original Hamiltonian and the bond model Hamiltonian in a finite volume $L$
with periodic boundary conditions 
are included in the spectrum of the Hamiltonian \eqref{chargeH} with some $L'\le L$, periodic boundary conditions, and an
appropriately chosen $\theta$. The special values that occur are given by $\theta=2\pi I/L^2$, where $I=0,\dots,L-1$.

In this work, we did not prove rigorously all the steps leading to this proposition, but the arguments could be
made precise by introducing wave functions for the spin part with appropriate periodic boundary conditions \cite{sajat-foldeds}.

We checked the claim numerically in small volumes, in both the integrable and non-integrable cases, and we observed that
it indeed holds.

\subsection{Integrability breaking}

We demonstrate here that the interaction term $\sum_j h_{ni}(j)$ in our original family of models breaks the
integrability. To this order we computed the level spacing statistics in finite volume. The original models have large
degeneracies, therefore it is more efficient to compute the energy levels from the family of models given by
\eqref{chargeH}. We consider only the case $\theta=0$, but this is enough to demonstrate the breaking of integrability.

\begin{figure}[h]
  \centering
\minipage{0.48\textwidth}
\includegraphics[scale=0.4]{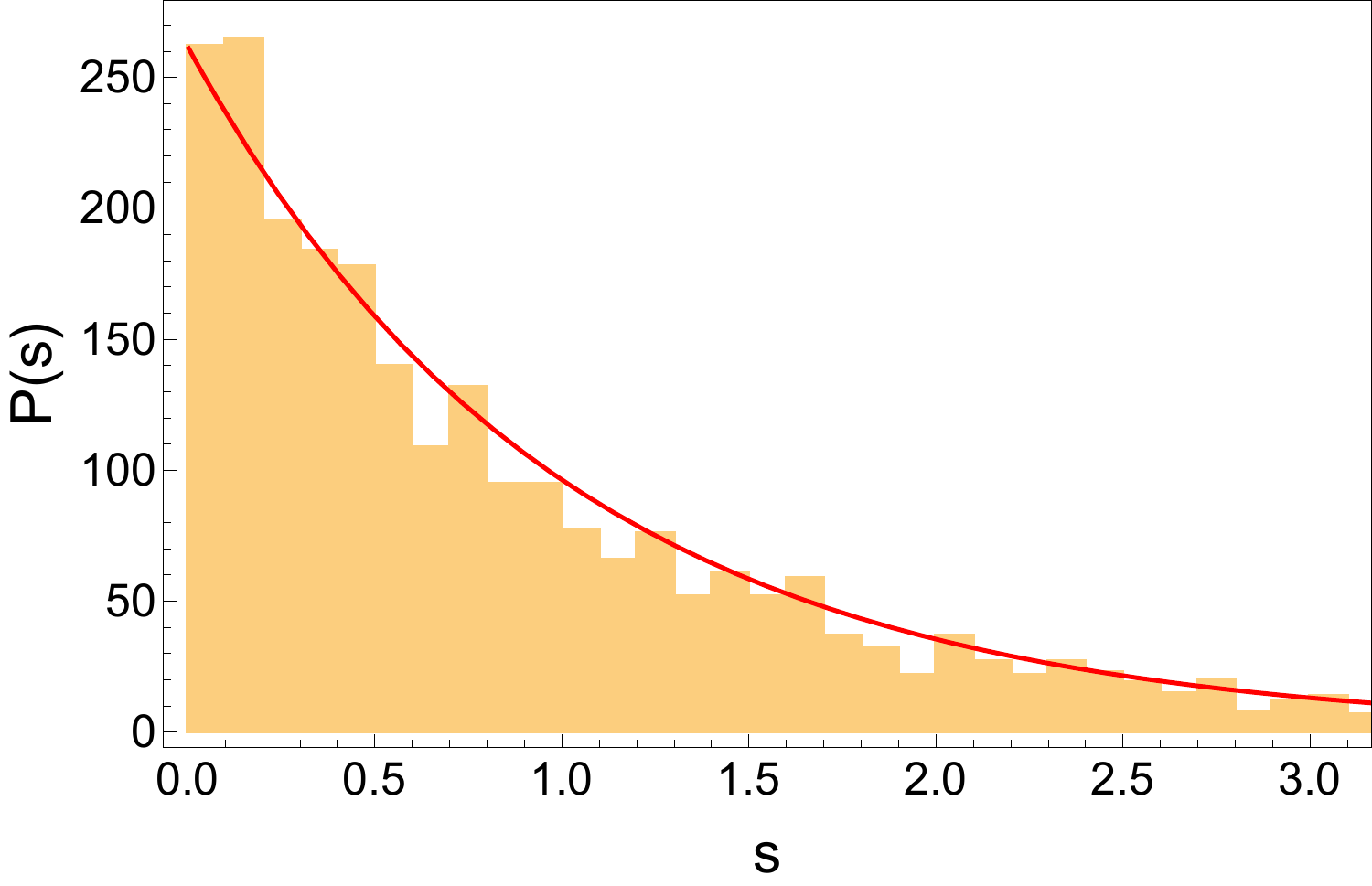}
\endminipage\hfill
\minipage{0.48\textwidth}
\includegraphics[scale=0.4]{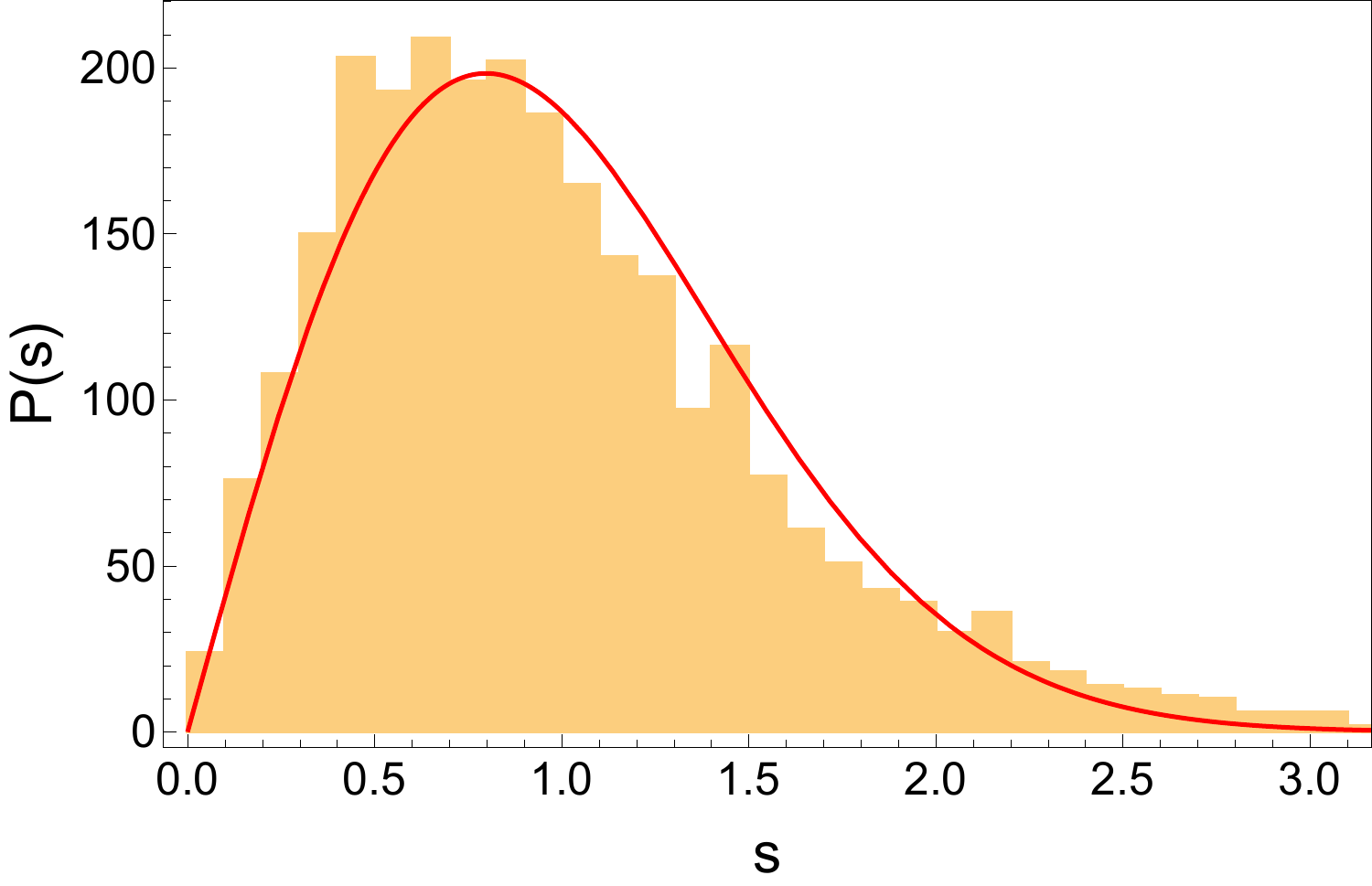}
\endminipage  
  \caption{Level spacing statistics in the integrable (left) and non-integrable (right) cases.}
  \label{fig:levels}
\end{figure}

The level spacing statistics of a given Hamiltonian $H$ with eigenvalues $\epsilon_i$ is obtained by taking the differences $S_i=\epsilon_{i+1}-\epsilon_i$ of the increasingly ordered eigen-energies $\epsilon_1\leq\epsilon_2\leq\dots$. The level spacing distribution $P(s)$ is then defined as the distribution of the normalized level spacings $s_i=S_i/\overline{S}$, where $\overline{S}$ is the mean level spacing. According to random matrix theory, for a non-integrable Hamiltonian taken from the Gaussian orthogonal ensemble, in infinite volume $P(s)$ is described by the Wigner-Dyson distribution:
\begin{equation}
P_{ni}(s)=\frac{\pi}{2}se^{-\frac{\pi}{4}s^2}.
\end{equation}
On the other hand, for an integrable system $P(s)$ is given by the exponential distribution:
\begin{equation}
P_{i}(s)=e^{-s}.
\end{equation}
Since the models defined by \eqref{chargeH} possess trivial symmetries, we compute the eigenvalues in the invariant subspace of zero total momentum, even spatial parity and fixed value of the $z$-component of the total spin, with $S_z=2$. (Similar results can be obtained for other values of $S_z$, however for $S_z=0$, there is an additional spin-flip symmetry.) Furthermore, we only consider energy levels from the middle third of the spectrum, because the structure of quasi-particle excitations coming from the ends of the spectrum causes deviations from the random matrix predictions. The level spacing distributions for $L=20$ and $\Delta=0.5$ are presented in Fig. \ref{fig:levels}, with $\kappa=0$ on the left (integrable case) and $\kappa=2$ on the right (non-integrable case). The results are well described by the exponential and Wigner-Dyson distributions (red curves in Fig. \ref{fig:levels}), with the total normalization of the distributions being the only fitting parameter. The small deviations from the theoretical predictions are caused by the finite size of the system, however the change in $P(s)$ caused by a non-zero $\kappa$ clearly demonstrates the integrability breaking effect of the term $h_{ni}(j)$.

\section{Construction of the MPO symmetries}

\label{sec:MPO}

In order to find the MPO's that perform the desired symmetry operations, we use the terminology and the methods of the work
\cite{mps-automatas}. The idea is to construct the MPO's in the form of a finite automaton. The states of the automaton
correspond to the basis states in the auxiliary space of the MPO's, while 
the transitions from one state to another describe the matrix elements (including the diagonal ones).  

We start with the MPO's for the deformed XXC models. In these cases the desired MPO's can be found directly, just by
using the observations about the spin-charge separation. Then we re-interpret these MPO's as an automaton.

Afterwards we construct new MPO's in the bond models,
and eventually in the 
original family of models. It is important that {\it we do not directly transform the MPO's of the XXC model}. Instead we
construct new MPO's that implement the action of the original MPO's together with the transformation rules between the
models. Therefore, there is no 
immediate connection between the dimensions of the auxiliary spaces. In the XXC model all MPO's have bond dimension 2.
The diagonal MPO's of the XXC model are realized in the original models with bond dimension 8, whereas the off-diagonal
ones will have bond dimension 10.

We use two different graphical notations for the MPO's. The first notation is the standard graphical representation of
the MPO as a tensor network, whereas the second one is a representation as an automaton.

In the first representation the fundamental tensor $\La_{a,j}$ is represented as a four leg object. We recall that
 $\La_{a,j}$ was introduced as a linear operator acting on $V_a\otimes V_j$, where $V_a$ is the $D$-dimensional auxiliary
space, and $V_j$ is the physical space at site $j$, which is 2 dimensional in the original and the bond models, and 3
dimensional in the XXC models.

\begin{figure}[h]
  \centering
  \includegraphics[scale=0.4]{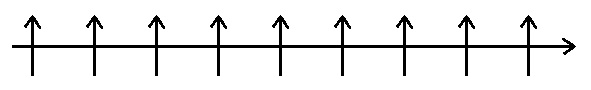}
  \caption{Graphical representation of an MPO.}
  \label{fig:mpo}
\end{figure}

The fundamental tensor $\La$ can always be expanded as
\begin{equation}
   \La=\sum_{a,b}\sum_{c,d} \La_{ab,cd} \ketbra{a}{b} \otimes  \ketbra{c}{d}.
\end{equation}
Here $\La_{ab,cd}$ are the coefficients for the local configurations of the four leg tensor.
In our notations the auxiliary space corresponds to the horizontal, whereas the
physical space to the vertical direction. The indices $a, b$ stand for the basis states in the auxiliary space, and $c,
d$ for the basis states in the physical space. The MPO is then represented in Fig. \ref{fig:mpo}; each crossing 
corresponds to the action of the linear operator $\La$, or alternatively, to the insertion of the four leg tensor into
the tensor network.

\subsection{MPO's for the XXC models}

For the XXC models we construct two families of MPO's with very different
physical meaning. The first family is diagonal, therefore they form a commutative symmetry algebra. The
second family is off-diagonal, and they change the spin part of the wave function.

In both cases we have $D=2$ and the operator $\La$ has
the restricted form given by eq. (9) in the main text.
The crucial property is that if the incoming or outgoing
state in the physical direction is a $\ket{0}$, then the action of the tensor becomes identical in both directions. This
has two implications:
\begin{enumerate}
\item  The resulting MPO acts as the identity on the charge degrees of freedom: the local states $\ket{0}$ are never moved.
\item The action of the MPO in the spin degrees of freedom is completely independent of the charge degrees of
  freedom. This happens because each time the MPO encounters a $\ket{0}$ state, the ``information stored'' in the
  auxiliary space gets copied.
\end{enumerate}

These properties can be formulated alternatively by considering the spin-charge separation given by formula
\eqref{statesep}.  All our MPO's are such that {\it they act only on the spin part of the wave functions, given by
$\psi_{a_1,a_2,\dots,a_N}$, while leaving the charge part invariant}.  This is in contrast with the XXC Hamiltonians, which
act only on the charge part, leaving the spin part invariant.

\begin{figure}[h]
    \centering
    \includegraphics[scale=0.2]{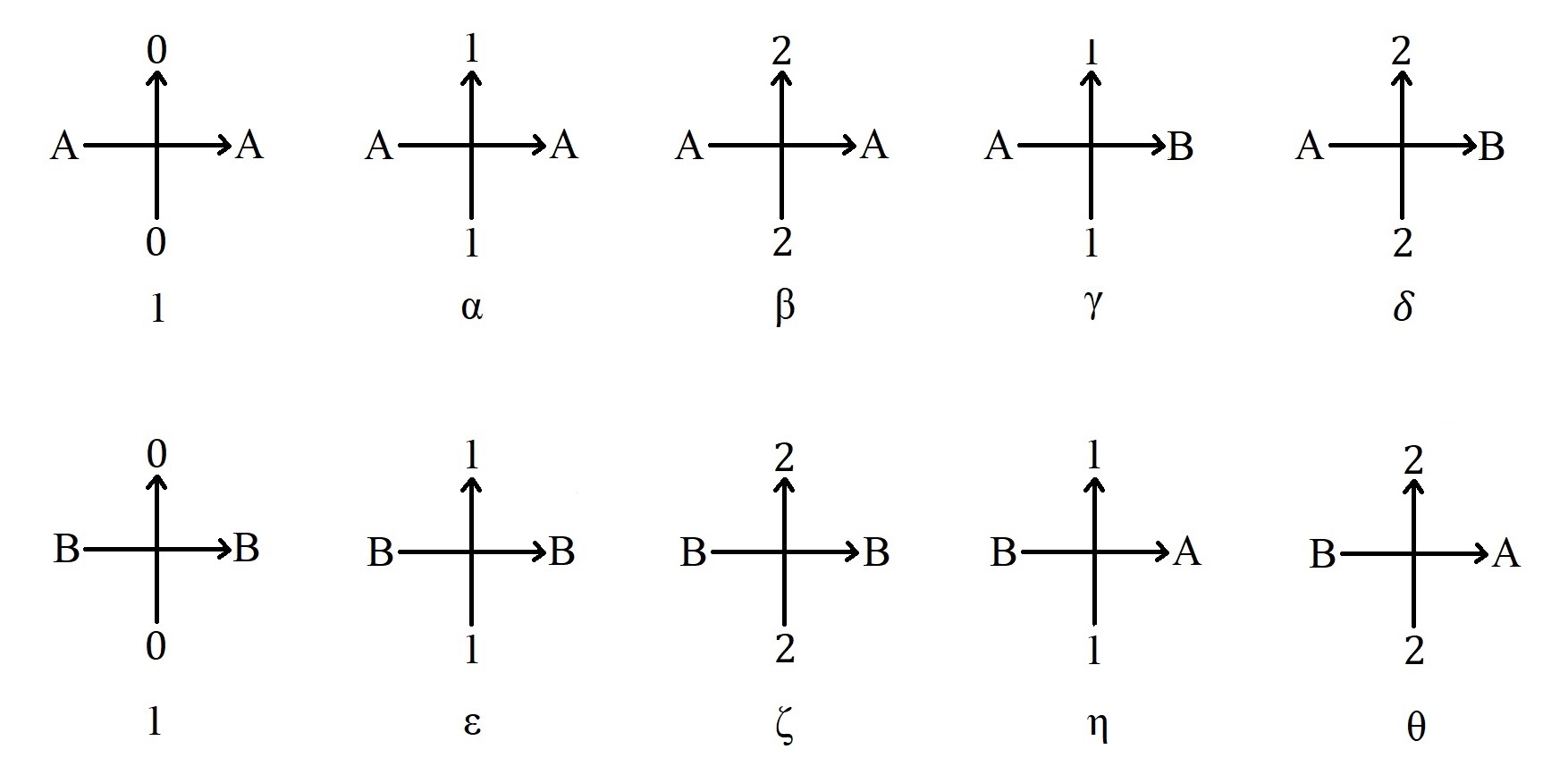}
    \caption{Tensor components of $\La$ in the XXC models, for the diagonal MPO's.}
    \label{fig:ladiag}
\end{figure}

\begin{figure}[h]
    \centering
    \includegraphics[scale=0.2]{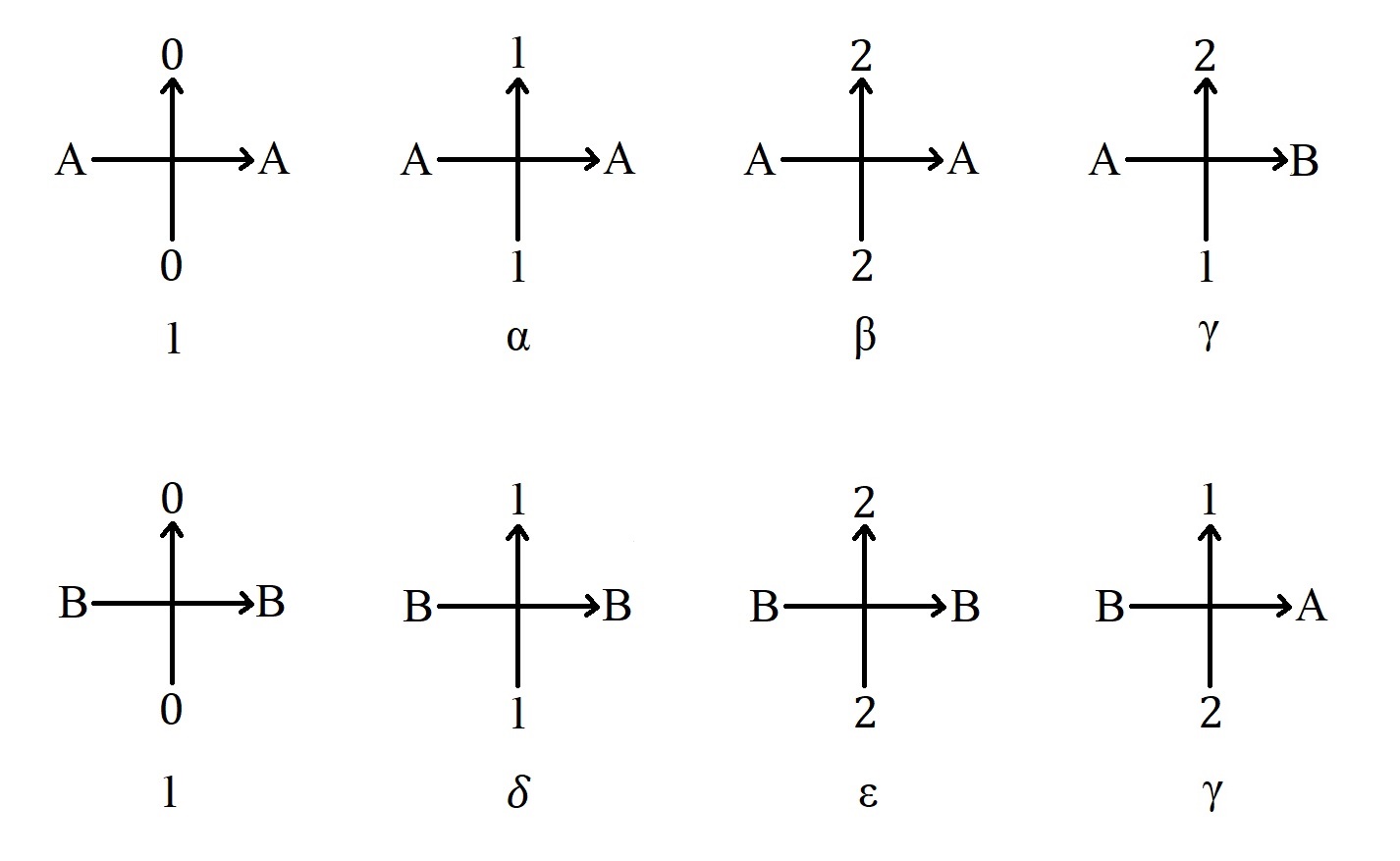}
    \caption{Tensor components of $\La$ in the XXC models, for the off-diagonal MPO's.}
    \label{fig:laoffdiag}
\end{figure}

Let us now consider the two concrete families of MPO's, given by eqs. (10) and (11) in the main text. 
The non-zero matrix elements are depicted also in Figs \ref{fig:ladiag}
and \ref{fig:laoffdiag}. Here the two basis states in the auxiliary space are denoted as $\ket{A}$ and $\ket{B}$.

In the case of the diagonal MPO's, the action of $\La$ is always diagonal in the physical space, in the given
basis. However, we allow for arbitrary transition matrix elements in the auxiliary space. Here in the Supplemental
Materials we depicted all possible matrix elements in our Figures. Nevertheless we note that the number of independent
parameters can be reduced by a similarity transformation in the auxiliary space, diagonalizing one of the non-vanishing
$F$ matrices. Choosing to diagonalize $F^{2,2}$ (and choosing an appropriate gauge via diagonal similarity
transformations) we obtain formula (10) of the main text. MPO's where neither matrix is diagonalizable are not important
for our main conclusions, therefore we do not investigate those cases in detail.

In the case of the off-diagonal MPO's, the fundamental tensor has 8 non-zero elements, and two of them correspond to the
identical action on the local physical states $\ket{0}$. The 6 non-trivial components coincide with those of the famous
six-vertex model \cite{Baxter-books}. This is an integrable model of two dimensional statistical physics, however, we do
not use its integrability in this work. The only important property is that the transfer matrix conserves the total
number of $\ket{1}$ and $\ket{2}$ states, and this is ensured locally, by spin preservation in the physical and
auxiliary spaces combined. Note that we chose the coefficients of the spin-flip components to be equal: this can always
be achieved by a diagonal similarity transformation in the auxiliary space, unless one of them is zero. 

\begin{figure}[h]
  \centering
\minipage{0.3\textwidth}
\includegraphics[scale=0.4]{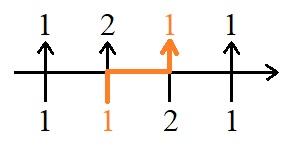}
\endminipage\hfill
\minipage{0.34\textwidth}
\includegraphics[scale=0.4]{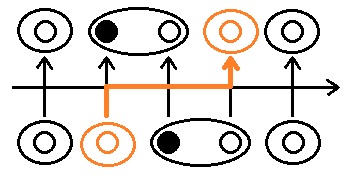}
\endminipage\hfill
\minipage{0.35\textwidth}
\includegraphics[scale=0.4]{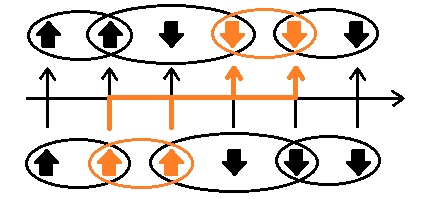}
\endminipage\hfill
\caption{An example for a rearrangement process in the computational basis. The action of the off-diagonal MPO's results
in linear combinations of such processes. The first picture shows the rearrangement in the XXC models, the second one
shows the same process translated to the bond model, finally the third picture shows the same process translated to the
original family of models. Note that the total magnetization is changed in  the final version of the process, which is
an effect of the displacement of a domain wall.}
  \label{fig:rearr1}
\end{figure}

For our purposes the action of the resulting MPO's is most important: These MPO's 
rearrange the spin pattern, and this is achieved in a very special and controlled way. Let us describe this
process. Within spin space (i.e. for the spin part of the wave function) we regard the states $\ket{1}$ as the vacuum,
and the states $\ket{2}$ as 
excitations. Accordingly, we regard the states $\ket{A}$ and $\ket{B}$ as the vacuum and excited state of the auxiliary
space. The MPO can displace excitations, and this happens as the result of a sequence of local steps. First, the
excitation is moved  from the physical space to the auxiliary space. Afterwards, the excitation might be transported
along a few sites. Finally, the excitation is moved from the auxiliary space to an other physical space. In this process
the auxiliary space is used as a ``memory storage''. It is important, that due to the small dimension of the auxiliary
space it can only store one excitation at a given time. Therefore, the displacements of the excitations on the physical
spin wave function happen only one by one. As an example see the first picture in Fig. \ref{fig:rearr1}.

Now we interpret these MPO's as an automaton. The two basis states of the auxiliary space will be interpreted as the two
states $A$ and $B$ of the automaton. The action of the operator $\La$ is represented as a sum of different transitions
between the states of the automaton, associated with the action of local operators on the local physical spaces. Acting
with the complete MPO (summation over internal states in the tensor network) is seen as a final effect of running the
automaton. 

\begin{figure}[h]
  \centering
  \minipage{0.5\textwidth}
    \includegraphics[scale=0.18]{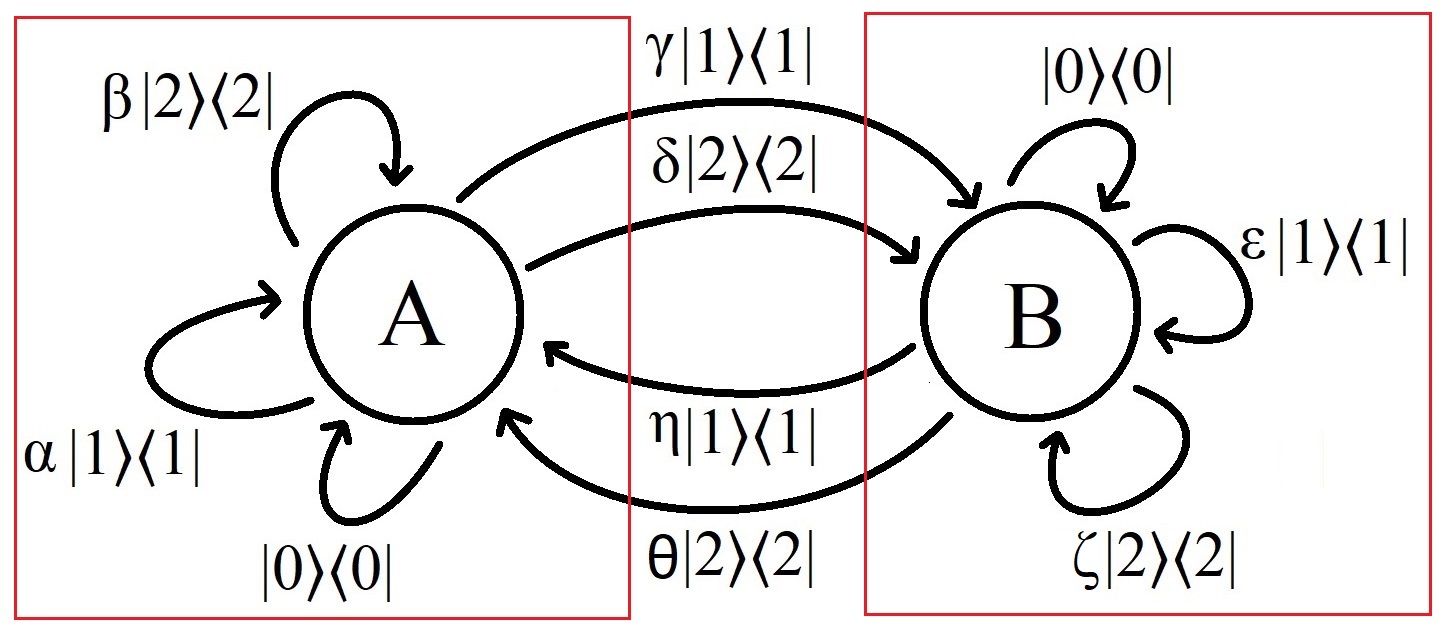}
\endminipage\hfill
\minipage{0.5\textwidth}
 \includegraphics[scale=0.18]{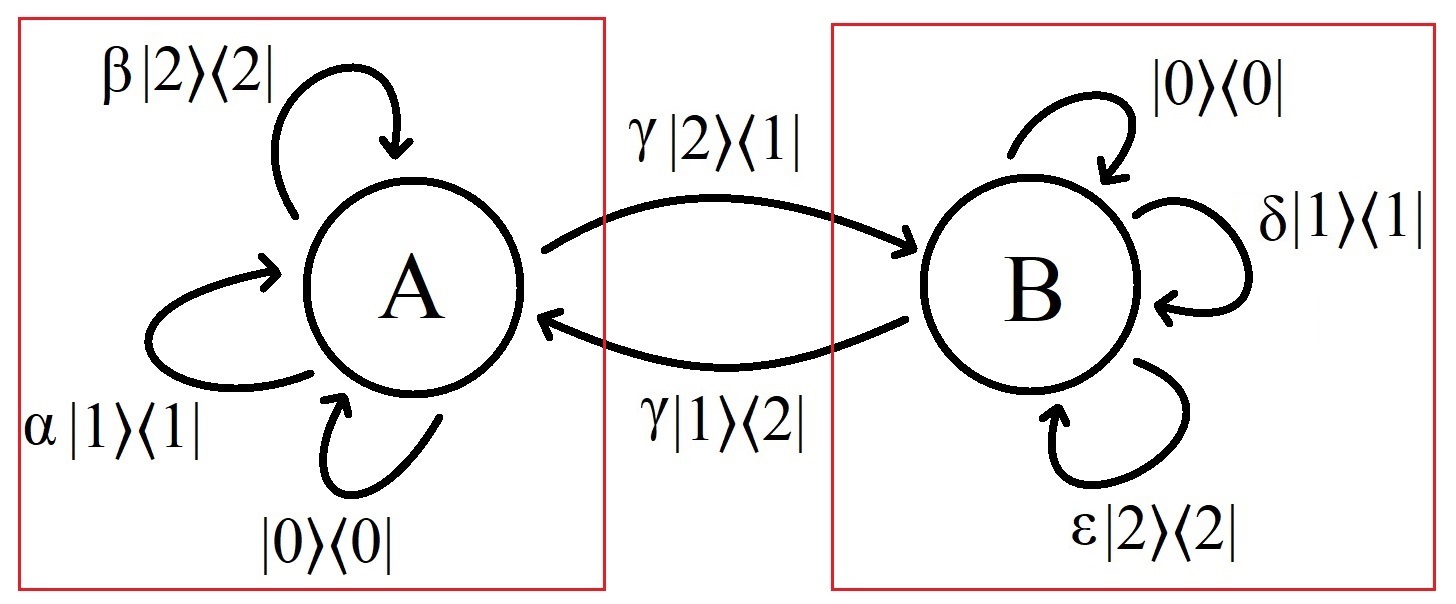}
\endminipage\hfill
    \caption{Automata for the MPO's in the deformed XXC model. Left: diagonal case, right: off-diagonal case. Both
      automata have two internal states $A$ and $B$, 
      corresponding to the two basis states in the auxiliary space. Arrows denote transitions with respect to auxiliary
      space, with the attached operator that acts on the physical space. }
    \label{fig:mm}
\end{figure}

In this case the non-zero elements of $\La$ are already given, and they are translated into the automata in a direct
way. The resulting two automata are depicted in Fig. \ref{fig:mm}.

\begin{figure}[h]
  \centering
\minipage{0.45\textwidth}
\includegraphics[scale=0.4]{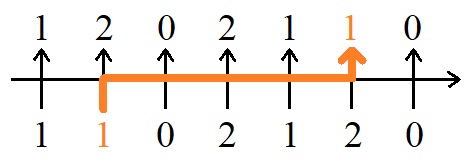}
\endminipage\hfill
\minipage{0.54\textwidth}
\includegraphics[scale=0.4]{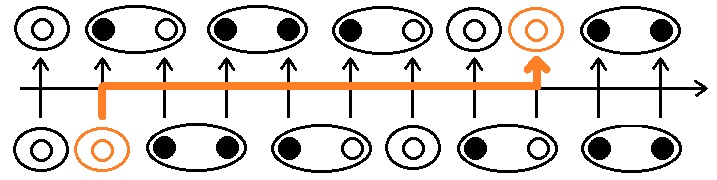}
\endminipage\hfill

\includegraphics[scale=0.4]{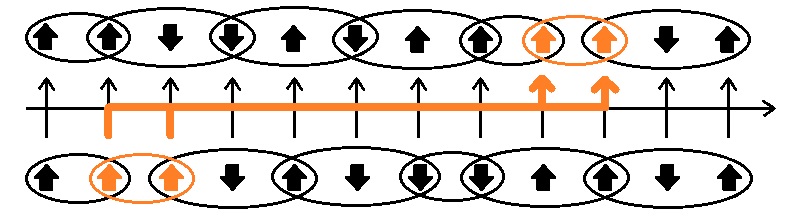}

\caption{A second example for a rearrangement process, showing all three versions. Note that the process is transparent
  in the XXC model, however, it gets more difficult to interpret in the original family of models. }
  \label{fig:rearr2}
\end{figure}

\subsection{MPO's for the bond models}

Now we construct the MPO's for the models given by \eqref{Hbond}. The idea is to construct two families of MPO's which
perform the same action as those in the XXC models. However, in these cases it is not immediately evident how to 
find the local tensors for the MPO's. The reason for this is that the transformation between the bond model and the XXC
models is very non-local, therefore the spin-charge separation is not transparent. Eventually, we intend to build MPO's
that leave the vacuum states invariant, and modify only the ``spin part'' of the wave functions. However, in the bond
model the vacuum states are given by the hard rods $\ket{\bullet\bullet}$, whereas the two different excitations are the
local states $\ket{\circ}$ and $\ket{\bullet\circ}$. The three different possibilities do not have equal length, and
this causes a considerable complication.

The main strategy behind our construction is the following. We build an automata, that ``reads'' the incoming sequence
of $\bullet$ and $\circ$ characters
(which corresponds to a given state in the computational basis), and it also ``translates'' it into a sequence of
characters 0, 1 and 2. Furthermore, the automata performs the same steps on this sequence, as in the case of the
XXC models. Finally, the automata writes the outcome of its action, once again ``translated'' into a sequence of
$\bullet$ and $\circ$ characters. The interpretation of the sequences is possible only if the automaton has an internal
``memory'', therefore the bond dimension needs to be bigger than in the XXC case. It turns out that we need to have an
auxiliary space of dimension 4 and 5, in the diagonal and off-diagonal 
cases, respectively.

The off-diagonal MPO's perform a rearrangement of different local configurations, for example see the processes depicted in
Figs. \ref{fig:rearr1} and \ref{fig:rearr2}. However, in the bond model it is the configurations $\ket{\circ}$ and
$\ket{\bullet\circ}$ which are 
exchanged (see the second pictures in both Figures). As an effect of the exchange the segment of the sequence which lies
between the two exchange points gets shifted by one site. 
An automaton with a finite number of internal states is enough
for our purposes, because the off-diagonal MPO's of the 
XXC model are such that such replacements happen only one at a time, therefore the maximum shift is only by one
site. Therefore, it is not required to have an infinite ``memory'' in the auxiliary space. 
 
\begin{figure}[h]
  \centering
   \minipage{0.5\textwidth}
    \includegraphics[scale=0.18]{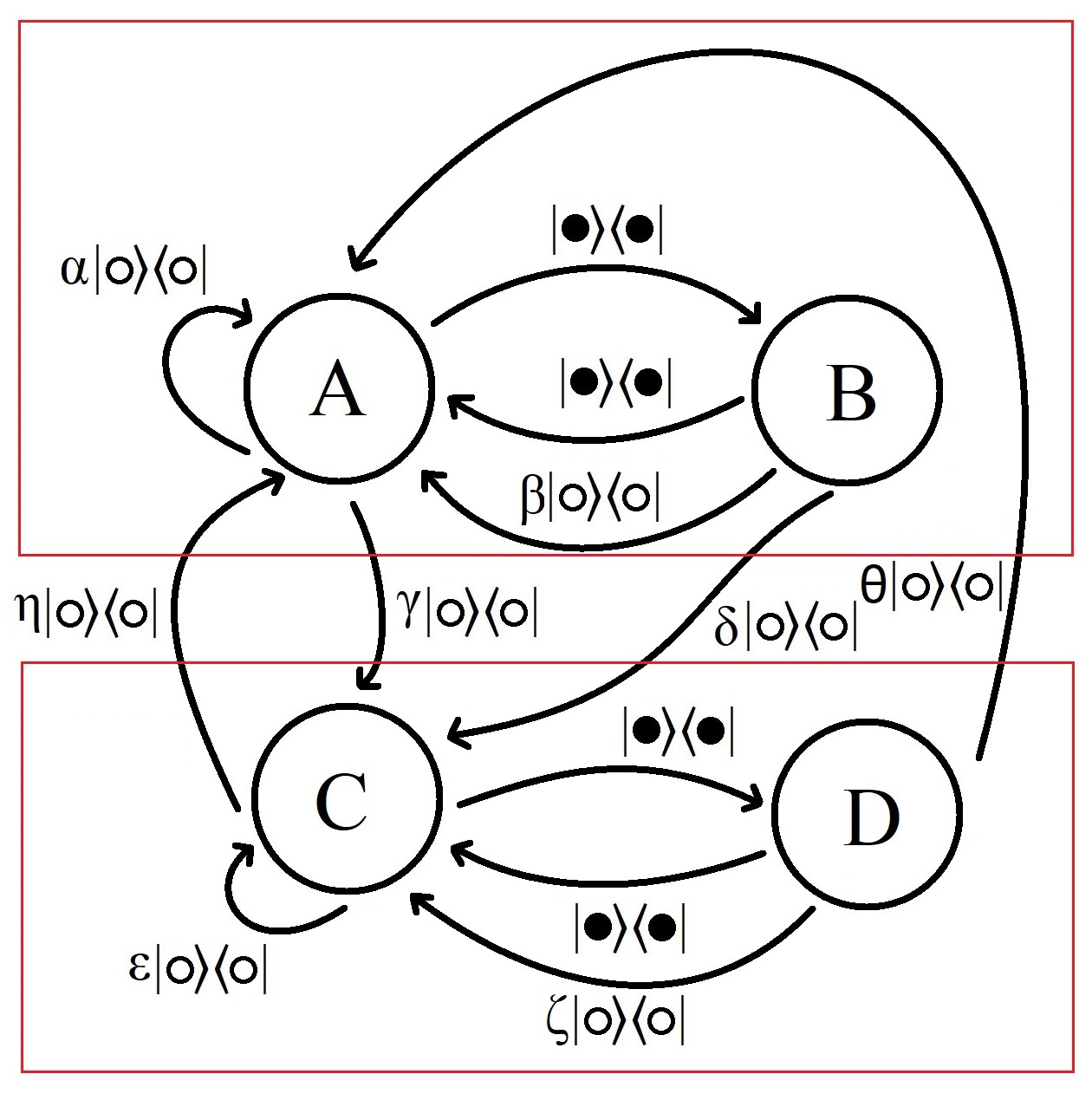}
    \endminipage\hfill
     \minipage{0.5\textwidth}
 \includegraphics[scale=0.18]{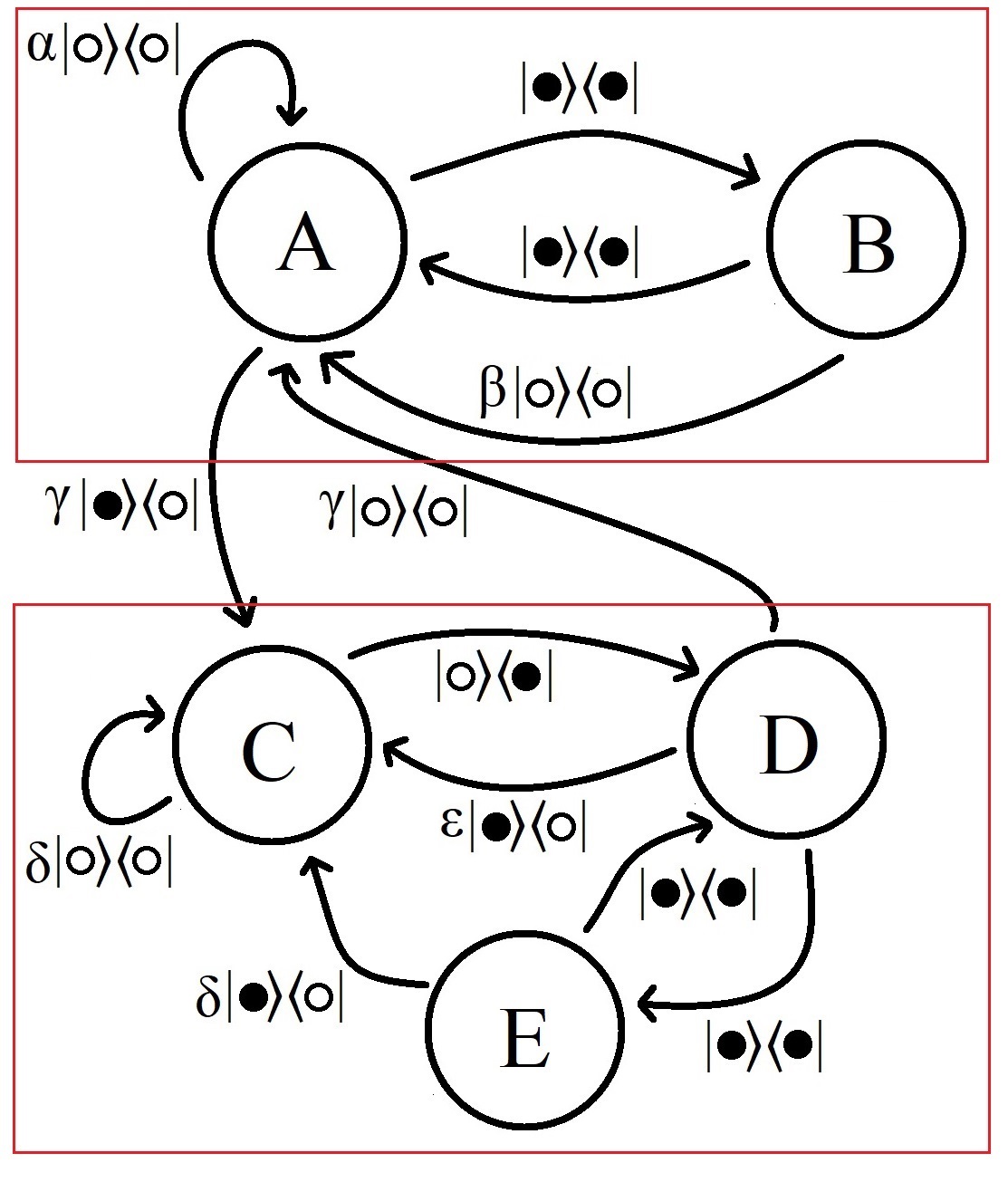}
\endminipage\hfill
       \caption{Automata for the MPO's in the bond model. Left: diagonal case, right: off-diagonal case.}
    \label{fig:bond}
  \end{figure}

The automaton for the diagonal MPO is shown in Fig. \ref{fig:bond} on the left. The red rectangles contain the same
units as in Fig. \ref{fig:mm}, but now two automaton states are necessary for each block due to the multi-site
configurations. Let us now describe how the MPO ``reads'' and ``writes'' the configurations.
The state $\ket{1}$ is recognized right away upon reading $\ket{\circ}$ in the automaton state $A\ (C)$, but further
information is needed after the local state $\ket{\bullet}$, hence the transition to automaton state $B\ (D)$. The
return to the origin takes place as the next site decides if we have $\ket{0}$ or $\ket{2}$.  Transition between the
blocks is possible after the successful recognition of either a state $\ket{1}$ or $\ket{2}$. We apply the same constant
factors as in the case of the XXC model. The fundamental tensor $\La$ of the MPO can easily be constructed from the automaton: states correspond to MPO dimensions and transitions between them to matrix elements. For the diagonal MPO we obtain
\begin{equation}
\La=    \begin{bmatrix}[cc|cc]
        \alpha & 0 & \gamma & 0 \\
        \beta & 0 & \delta & 0 \\ \hline
        \eta & 0 & \varepsilon & 0 \\
        \theta & 0 & \zeta & 0 \\
    \end{bmatrix}\otimes
    \ketbra{\circ}{\circ} + 
    \begin{bmatrix}[cc|cc]
        0 & 1 & 0 & 0 \\
        1 & 0 & 0 & 0 \\ \hline
        0 & 0 & 0 & 1 \\
        0 & 0 & 1 & 0 \\
    \end{bmatrix}\otimes
    \ketbra{\bullet}{\bullet},
\end{equation}
which can be factorized as
\begin{equation}
\La=    \Bigg(
    \begin{bmatrix}
        \alpha & \gamma \\
        \eta & \varepsilon \\
    \end{bmatrix}
    \otimes
    \begin{bmatrix}
        1 & 0 \\
        0 & 0 \\
    \end{bmatrix}
    +
    \begin{bmatrix}
        \beta & \delta \\
        \theta & \zeta \\
    \end{bmatrix}
    \otimes
    \begin{bmatrix}
        0 & 0 \\
        1 & 0 \\
    \end{bmatrix}
    \Bigg) \otimes \ketbra{\circ}{\circ}
    +
    \begin{bmatrix}
        1 & 0 \\
        0 & 1 \\
    \end{bmatrix}
    \otimes
    \begin{bmatrix}
        0 & 1 \\
        1 & 0 \\
    \end{bmatrix}\otimes
    \ketbra{\bullet}{\bullet}.
\end{equation}
The origin of this factorization property in the automaton is the presence of the identical upper and lower
parts. Similar to the XXC case, this
form allows us to reduce the number of free parameters by diagonalizing one of the matrices in the first vector
space. After that, we can choose the off-diagonal elements of the other matrix to be equal without the loss of
generality. Thus the final number of parameters is five instead of eight. Once again, we dismiss those cases where
neither matrix is diagonalizable, as they are not important for our conclusions.

In the case of the off-diagonal MPO the automaton is shown in Fig. \ref{fig:bond} on the right.  Now there are five internal states.
The upper block is identical to the one described above. Transition to the more complicated lower block can be triggered
by reading a state $\ket{\circ}$, that is, identifying the incoming state $\ket{1}$. The MPO can perform the exchange $\ket{1}
\rightarrow \ket{2}$, but in the bond model this introduces a volume change ($\ket{\circ} \rightarrow
\ket{\bullet\circ}$) resulting in a shift for all following sites. Therefore the task of the lower block on each site is
to read the new state while printing the previous one (starting with an extra $\ket{\circ}$ from the initial
exchange). For the implementation we need one extra automaton state in this block, thus we have five in total. One arrives in
automaton state $C$ after reading $\ket{\circ}$, in $D$ after an odd number of consecutive $\ket{\bullet}$'s and in $E$
after an even number of consecutive $\ket{\bullet}$'s. The transition from the lower block to the upper one takes place
upon reading $\ket{\bullet\circ}$ but only printing $\ket{\circ}$ thus performing the exchange
$\ket{2}\rightarrow\ket{1}$. We get 
\begin{equation}
\La=    \begin{bmatrix}
        \delta & 0 & 0 & 0 & 0 \\
        \varepsilon & 0 & 0 & 0 & 0 \\
        0 & 0 & \alpha & 0 & 0 \\
        \gamma & 0 & 0 & 0 & 0 \\
        0 & 0 & 0 & 0 & 0
    \end{bmatrix}\otimes
    \ketbra{\circ}{\circ} + 
    \begin{bmatrix}
        0 & 0 & \gamma & 0 & 0 \\
        0 & 0 & 0 & 0 & 0 \\
        0 & 0 & 0 & 0 & 0 \\
        0 & 0 & \beta & 0 & 0 \\
        0 & 0 & \alpha & 0 & 0
    \end{bmatrix}\otimes
    \ketbra{\circ}{\bullet}
    + 
    \begin{bmatrix}
        0 & 0 & 0 & 0 & 0 \\
        0 & 0 & 0 & 0 & 0 \\
        0 & 0 & 0 & 1 & 0 \\
        0 & 0 & 0 & 0 & 0 \\
        0 & 0 & 0 & 0 & 0
    \end{bmatrix}\otimes
    \ketbra{\bullet}{\circ} + 
    \begin{bmatrix}
        0 & 1 & 0 & 0 & 0 \\
        1 & 0 & 0 & 0 & 0 \\
        0 & 0 & 0 & 0 & 0 \\
        0 & 0 & 0 & 0 & 1 \\
        0 & 0 & 0 & 1 & 0
    \end{bmatrix}\otimes
    \ketbra{\bullet}{\bullet}.
\end{equation}
 
We remark that another construction of the off-diagonal MPO is possible where the exchange $\ket{2} \rightarrow \ket{1}$ introduces the shift which has the opposite direction (left instead of right). Formally this can be obtained by the following exchanges
\begin{align}
    \alpha &\longleftrightarrow \delta, \\
    \beta &\longleftrightarrow \varepsilon, \\
    \ketbra{\bullet}{\circ} &\longleftrightarrow \ketbra{\circ}{\bullet}.
\end{align}
However, the only difference between the action of the two MPO's is an overall one-site shift.

The alerted reader might wonder whether these MPO's are truly compatible with periodic boundary conditions. After all,
the rules for the automata were found by assuming that the automaton can read and immediately ``translate'' a given
sequence. However, there is the ambiguity of choosing a starting point for the assignment of particle and vacuum labels,
and also the
problem that a sequence ending with an occupied site does not yield a proper state in the XXC models. Fortunately, for
every state of 
the bond model there is exactly one way to assign the labels properly. As an effect, the MPO's will have a well defined
and correct action if we use periodic boundary conditions. Let us prove this statement in more detail.
 
In the bond model let us assume that there is at least one empty
site. Let us pick this site and start the labeling at the next site. Doing so will always result in a well-defined XXC
state, because the sequence ends 
with an empty site, and every proper labeling is like that.
On the other hand, if 
all the sites are occupied, then the state corresponds to the vacuum of the XXC model for an even number of sites. For an odd
number of sites we can not have all sites occupied since an odd number of spins can not alternate periodically. All this
leads to the conclusion that the correspondence between the states of the bond model and the XXC models is well defined,
and the only problem that might occur is that the ``translation procedure'' might need to be started from a site
different from the first site. This does not cause any problems if we use periodic MPO's.

\subsection{MPO's for the original family of models}

The states of the original model can be reconstructed from the states of the bond model only up to a global spin flip,
or equivalently, if we know the first spin of a given sequence.
This implies that in the original model  one needs an
extra ``memory bit'' in the automaton, which takes care of the extra information. In practice, however, the extra memory
bit is not decoupled, and one needs a careful investigation of the various possibilities, in order to arrive at the
final automaton and the corresponding MPO. Eventually one has to duplicate each element of the bond model automata
because the last spin can be either up or down upon arriving at an automaton state.

\begin{figure}[b]
    \centering
    \includegraphics[scale=0.15]{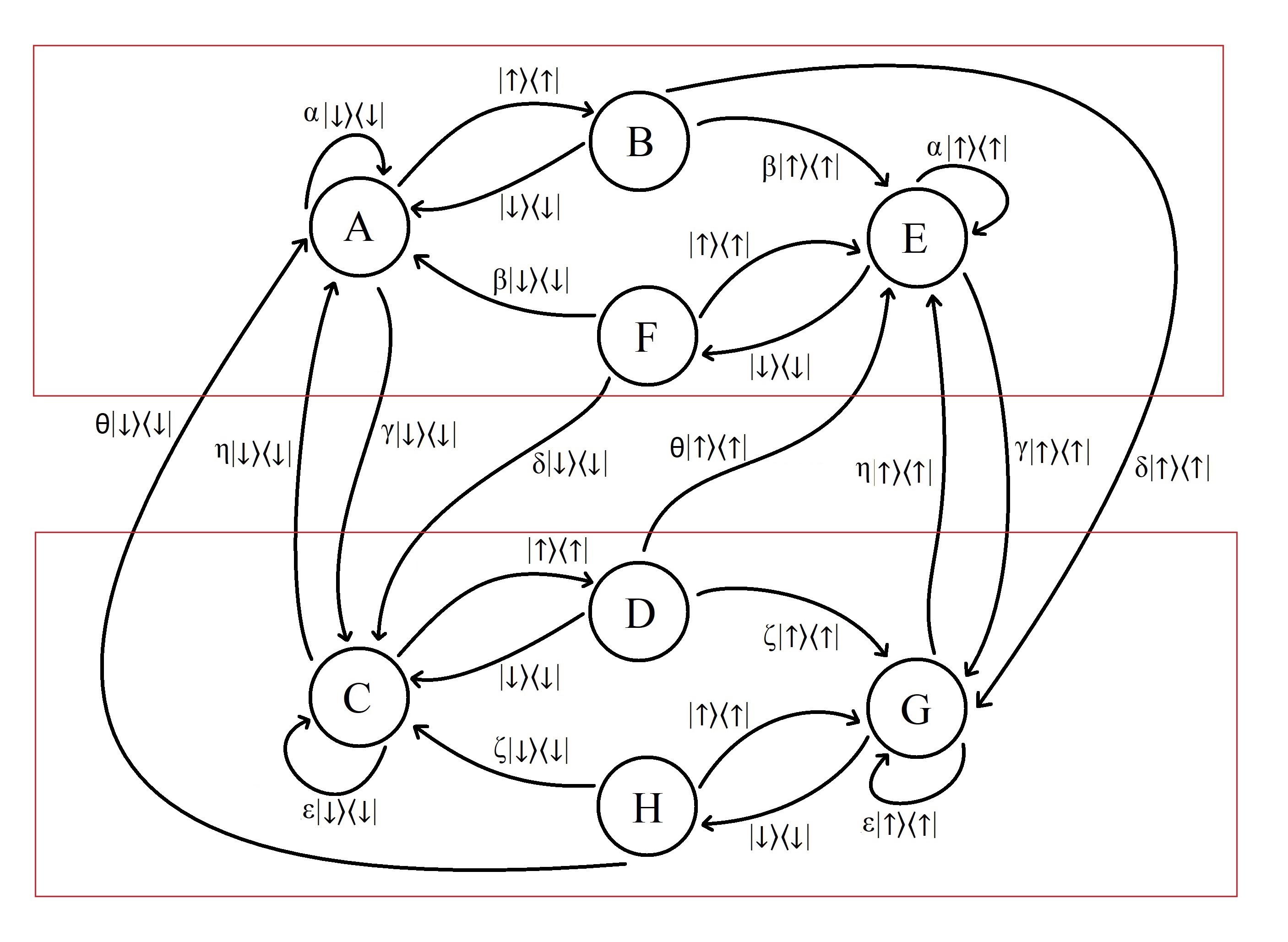}
    \caption{Automaton for the diagonal MPO in the original model.}
    \label{fig:diag-fxxz}
\end{figure}

The automaton of the diagonal MPO is shown in Fig. \ref{fig:diag-fxxz}. The elementary matrix can now be doubly
factorized: we have the two copies of the bond model automaton (corresponding to states A-D and E-H), each of them
having the ``upper'' and ``lower'' 
parts. Regarding the formulas for $\La$ here we only give 
the final form: 
\begin{equation}
    \begin{split}
\La=    \begin{bmatrix}
        \ketbra{\downarrow}{\downarrow} & 0 \\
        0 & \ketbra{\uparrow}{\uparrow} \\
    \end{bmatrix}
    &\otimes \Bigg( 
    \begin{bmatrix}
        \alpha & \gamma \\
        \eta & \varepsilon \\
    \end{bmatrix}
    \otimes
    \begin{bmatrix}
        1 & 0 \\
        0 & 0 \\
    \end{bmatrix}
    +
    \begin{bmatrix}
        1 & 0 \\
        0 & 1 \\
    \end{bmatrix}
    \otimes
    \begin{bmatrix}
        0 & 0 \\
        1 & 0 \\
    \end{bmatrix}
    \Bigg) \\
    &+
    \begin{bmatrix}
        \ketbra{\uparrow}{\uparrow} & 0 \\
        0 & \ketbra{\downarrow}{\downarrow} \\
    \end{bmatrix}
    \otimes
    \begin{bmatrix}
        1 & 0 \\
        0 & 1 \\
    \end{bmatrix}
    \otimes 
    \begin{bmatrix}
        0 & 1 \\
        0 & 0 \\
    \end{bmatrix}
    +
    \begin{bmatrix}
        0 & \ketbra{\uparrow}{\uparrow} \\
        \ketbra{\downarrow}{\downarrow} & 0 \\
    \end{bmatrix}
    \otimes
    \begin{bmatrix}
        \beta & \delta \\
        \theta & \zeta \\
    \end{bmatrix}
    \otimes
    \begin{bmatrix}
        0 & 0 \\
        1 & 0 \\
    \end{bmatrix}.
    \end{split}
\end{equation}
Here we used a mixed notation, where $\La$ is represented as a matrix of size $8 \times 8$ acting in the auxiliary
space, with matrix elements given by operators acting on the physical space. Separating the coefficients according to
the operators acting on the physical space one gets eq. (13) of the main text.

This representation also allows for the reduction of the number of parameters the same way as described for the bond
model automaton (now diagonalizing in the second vector space). 

\begin{figure}[b]
    \centering
    \includegraphics[scale=0.15]{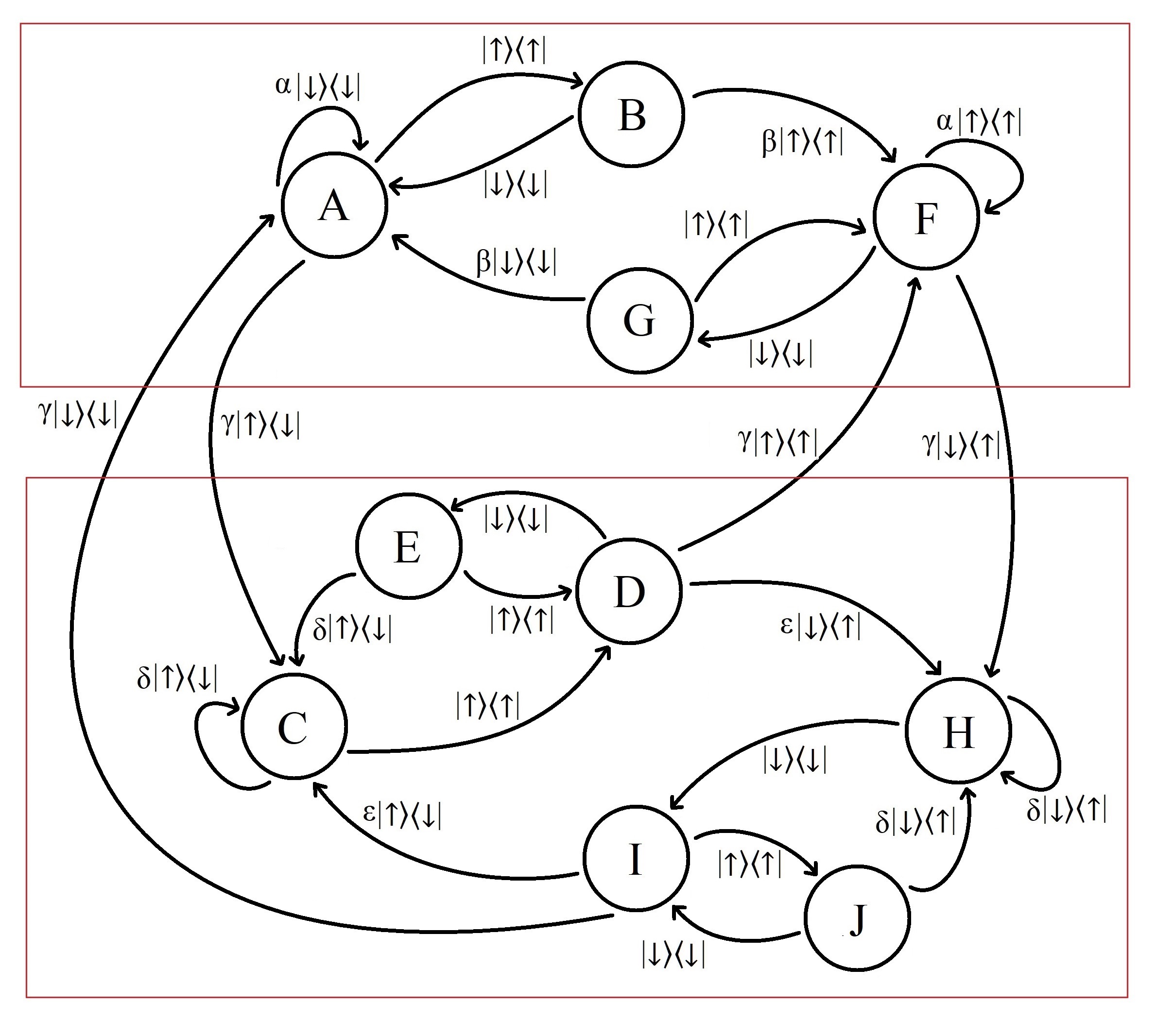}
    \caption{Automaton for the off-diagonal MPO in the original model.}
    \label{fig:exchange1-fxxz}
\end{figure}

Finally, the off-diagonal automaton is shown in Fig. \ref{fig:exchange1-fxxz}. Factorization of the elementary matrix
is once again possible due to the two copies of the bond model automaton, corresponding to states A-E and F-J. The final
result is 
\begin{equation}
    \begin{split}
\La=    \begin{bmatrix}
        \ketbra{\uparrow}{\uparrow} & 0 \\
        0 & \ketbra{\downarrow}{\downarrow} \\
    \end{bmatrix}
    \otimes 
    \begin{bmatrix}
        0 & 1 & 0 & 0 & 0 \\
        0 & 0 & 0 & 0 & 0 \\
        0 & 0 & 0 & 1 & 0 \\
        0 & 0 & 0 & 0 & 0 \\
        0 & 0 & 0 & 1 & 0 \\
    \end{bmatrix}
    +
    \begin{bmatrix}
        \ketbra{\downarrow}{\downarrow} & 0 \\
        0 & \ketbra{\uparrow}{\uparrow} \\
    \end{bmatrix}
    \otimes
    \begin{bmatrix}
        \alpha & 0 & 0 & 0 & 0 \\
        1 & 0 & 0 & 0 & 0 \\
        0 & 0 & 0 & 0 & 0 \\
        0 & 0 & 0 & 0 & 0 \\
        0 & 0 & 0 & 0 & 0 \\
    \end{bmatrix}
    +
    \begin{bmatrix}
        \ketbra{\uparrow}{\downarrow} & 0 \\
        0 & \ketbra{\downarrow}{\uparrow} \\
    \end{bmatrix}
    \otimes
    \begin{bmatrix}
        0 & 0 & \gamma & 0 & 0 \\
        0 & 0 & 0 & 0 & 0 \\
        0 & 0 & \delta & 0 & 0 \\
        0 & 0 & 0 & 0 & 0 \\
        0 & 0 & \delta & 0 & 0 \\
    \end{bmatrix}
    \\
    + 
    \begin{bmatrix}
        0 & \ketbra{\uparrow
        }{\uparrow} \\
        \ketbra{\downarrow}{\downarrow} & 0 \\
    \end{bmatrix}
    \otimes
    \begin{bmatrix}
        0 & 0 & 0 & 0 & 0 \\
        \beta & 0 & 0 & 0 & 0 \\
        0 & 0 & 0 & 0 & 0 \\
        \gamma & 0 & 0 & 0 & 0 \\
        0 & 0 & 0 & 0 & 0 \\
    \end{bmatrix}
    +
    \begin{bmatrix}
        0 & \ketbra{\downarrow}{\uparrow} \\
        \ketbra{\uparrow}{\downarrow} & 0 \\
    \end{bmatrix}
    \otimes
    \begin{bmatrix}
        0 & 0 & 0 & 0 & 0 \\
        0 & 0 & 0 & 0 & 0 \\
        0 & 0 & 0 & 0 & 0 \\
        0 & 0 & \varepsilon & 0 & 0 \\
        0 & 0 & 0 & 0 & 0 \\
    \end{bmatrix}.
    \end{split}
\end{equation}
Once again we used the mixed notation explained above. Separating the coefficients of the operators acting on the
physical space, and collecting the non-zero matrix elements one obtains eq. (14) of the main text.

\subsection{Classical vs. quantum fragmentation}

The diagonal MPO's belong to the commutant algebra: they commute with each term of the bond model
Hamiltonian \eqref{Hbond}. This is true already in the XXC model: there the MPO's are sensitive only to the spin part of
the wave function, and 
they do not change it, and each term in the Hamiltonian also keeps the spin part invariant. The diagonal MPO's in the
bond model preserve this commutativity, because they do not rearrange the particle positions, therefore there is no
effect of non-locality. The same holds true in the original formulation of our models.

In the deformed XXC models the off-diagonal MPO's also belong to the commutant algebra. Therefore, according to
\cite{fragm-commutant-1s} the deformed XXC models should be seen as having quantum fragmentation.

Remarkably, in the bond model the off-diagonal MPO's do not belong to the commutant algebra anymore.  They commute only
with the extensive Hamiltonian, but not 
with each term separately. This happens due to the non-local effects: The MPO's rearrange the particle content, leading
to a one-site shift of certain segments of the chain. See for example the second picture in Fig. \ref{fig:rearr2}. Such
shifts do not commute with the Hamiltonian densities that act inside of such segments, and this is the reason why these
MPO's do not belong to the commutant algebra. Following \cite{fragm-commutant-1s} this would imply that our models do
not have quantum fragmentation. However, we believe it is adequate to say that quantum fragmentation happens in our
models, because of the presence of the large family of non-commuting, off-diagonal symmetry operators, with low spatial
entanglement. 

\section{The \lowercase{i}TEBD algorithm} 

To demonstrate that the family of models treated in this work supports persistent oscillations, we numerically computed
the time evolution following a quantum quench, using the iTEBD algorithm \cite{vidal-itebd0s,vidal-itebd1as}.  We used the example code in \cite{pollmann-notess}
as a starting point and modified it to our purposes to simulate real-time evolution governed by a six-site Hamiltonian:
the state of the system is represented as a six-site translational invariant matrix product state (MPS). The
algorithm uses a first order Suzuki-Trotter decomposition for the time evolution operator. We initialized our system in
the state completely polarized in the $x$-direction, evolved it with a Trotter time step of $\delta t = 0.01$ and
calculated the expectation value of the operator $X$ as a function of time. To check the validity of our results, we
used several different maximal bond dimensions $\chi_{max}$. As the entanglement entropy grows in time, larger and
larger $\chi_{max}$ is needed to adequately approximate the state of the system, eventually limiting the time scale
reachable by the numerical method. The curves obtained with $\chi_{max}=500$ are presented in the main text, and they
support the theoretical predictions in the investigated time frame.

\end{document}